\documentclass[11pt]{article}
\pdfoutput=1

\usepackage{graphicx}
\usepackage{amsmath,amsfonts,amssymb,amsthm}
\usepackage{float,dsfont,url,color}
\usepackage{multirow}
\usepackage{ulem}
\newcommand{\bs}{\boldsymbol}
\usepackage[authoryear]{natbib}
\usepackage{authblk}

\newtheorem{mydef}{Definition}

\begin{document}

\title{Bayesian estimation of Differential Transcript Usage from RNA-seq data}
\author[1]{Panagiotis Papastamoulis\thanks{corresponding author: panagiotis.papastamoulis@manchester.ac.uk}}
\author[1]{Magnus Rattray}
\affil[1]{Division of Informatics, Imaging and Data Sciences\\Faculty of Biology, Medicine and Health\\University of Manchester, UK}

\renewcommand\Authands{ and }

\date{}

\maketitle

\begin{abstract}
Next generation sequencing allows the identification of genes consisting of differentially expressed transcripts, a term which usually refers to changes in the overall expression level. A specific type of differential expression is differential transcript usage (DTU) and targets changes in the relative within gene expression of a transcript. The contribution of this paper is to: (a) extend the use of cjBitSeq to the DTU context, a previously introduced Bayesian model which is originally designed for identifying changes in overall expression levels and (b) propose a Bayesian version of DRIMSeq, a frequentist model for inferring DTU. cjBitSeq is a read based model and performs fully Bayesian inference by MCMC sampling on the space of latent state of each transcript per gene. BayesDRIMSeq is a count based model and estimates the Bayes Factor of a DTU model against a null model using Laplace's approximation. The proposed models are benchmarked against the existing ones using a recent independent simulation study as well as a real RNA-seq dataset. Our results suggest that the Bayesian methods exhibit similar performance with DRIMSeq in terms of precision/recall but offer better calibration of False Discovery Rate.
\end{abstract}

\section{Introduction}

High throughput sequencing of cDNA (RNA-seq) \citep{mort} is an important tool to quantify transcript expression levels and to identify differences between different biological conditions. RNA-seq experiments produce a large number (millions) of short reads (nucleotide sequences) which are typically mapped to the genome or transcriptome. Expression quantification requires estimating the number of reads originating from each transcript in a given sample. Quantifying the transcriptome between different samples allows the identification of differentially expressed (DE) transcripts between them. However, certain difficulties complicate the inference procedure. In higher eukaryotes, most genes are spliced into alternative transcripts which share specific parts of their sequence (exons). Hence, a given short read typically aligns to different positions of the transcriptome and statistical models are often used to infer the origin probabilistically \citep{cuffdif, cuffdif2,rsem,isoEM,bitseq,casper,bitseqVB}.

Differential Transcript Expression (DTE) refers to the event where the overall relative expression of a transcript changes between two conditions. In this case, $\theta_k$ refers to the relative expression of transcript $k$; $k=1,\ldots,K$, with respect to the whole set of transcripts, with $\theta_k\geqslant 0$ and $\sum_{k=1}^{K} \theta_k= 1$. On the contrary, Differential Transcript Usage (DTU) refers to the event that the relative within gene abundance of a transcript changes between conditions. Consider a gene $g=1,\ldots,G$ with $K_g > 1$ transcripts. Then, the relative within gene transcript abundance is defined as $\theta_k^{(g)}=\frac{\theta_k}{\sum_{j \in g}\theta_j}$. Obviously, if a transcript belongs to a gene with $K_g = 1$ then it is always non-DTU. According to \cite{gonzalez2013transcriptome} the dominant transcripts within a gene are likely to be the main contributors to the proteome and switching events between them is a common scenario of gene modification between conditions.

\begin{figure}[t]
\centering
\begin{tabular}{c}
\includegraphics[scale=0.35]{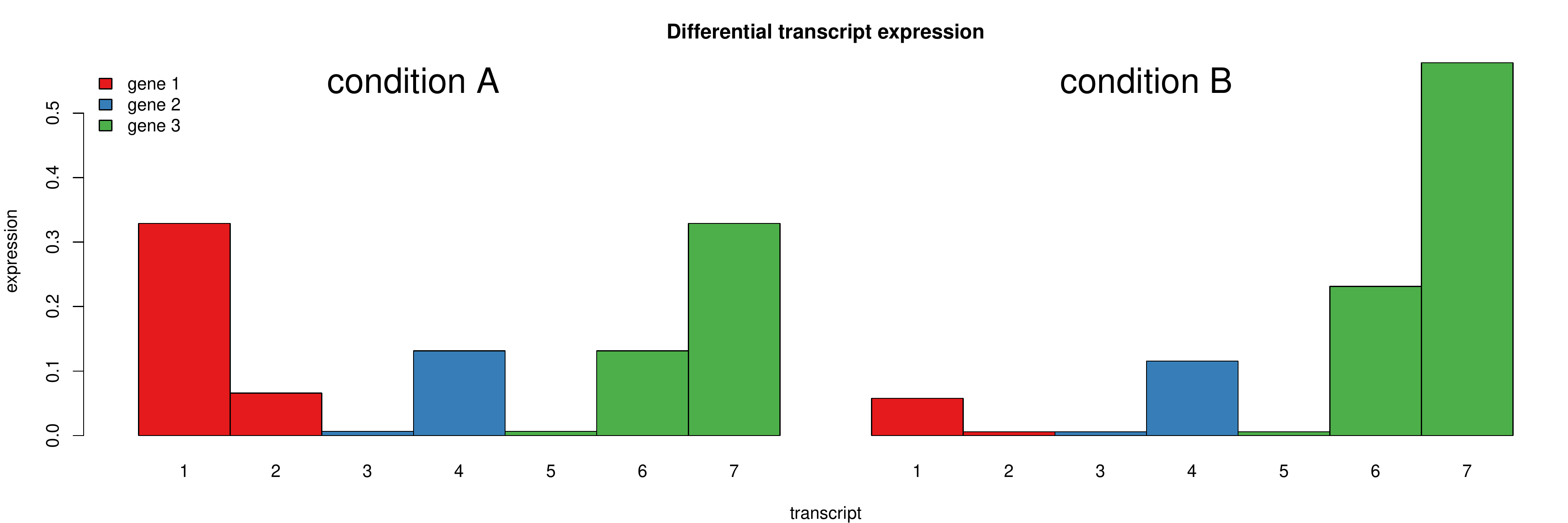} \\
\includegraphics[scale=0.35]{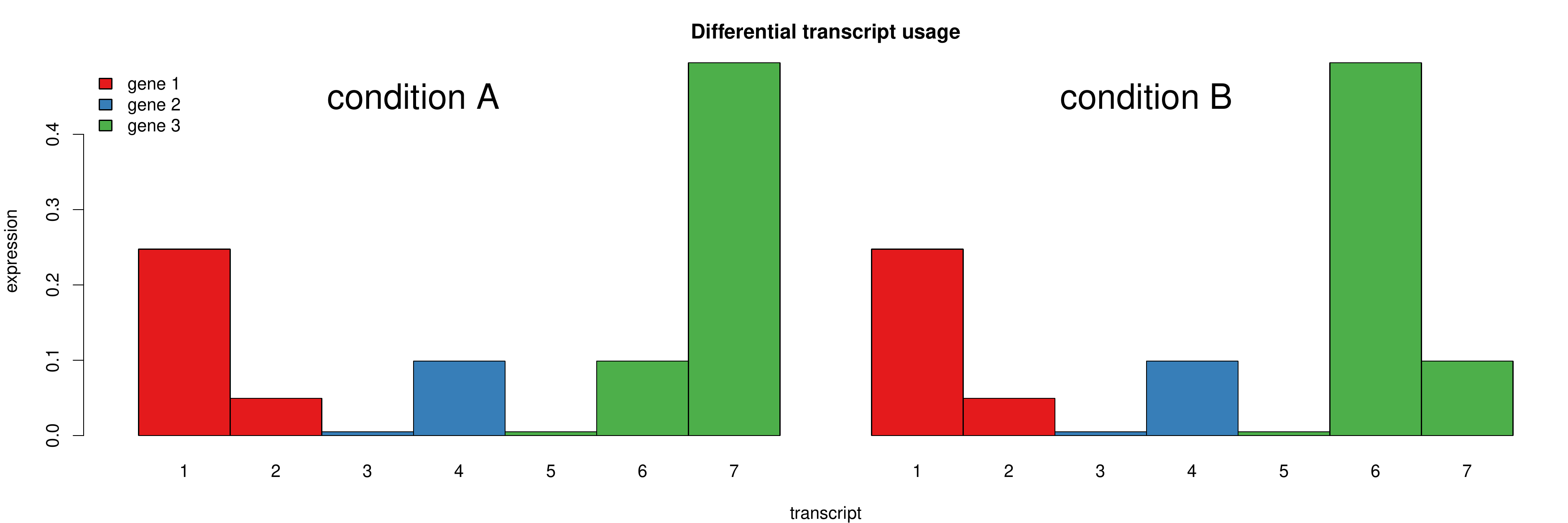}
\end{tabular}
\caption{Differential Transcript Expression (up) and Differential Transcript Usage (down).}\label{fig:dteVSdtu}
\end{figure}

Figure \ref{fig:dteVSdtu} illustrates the differences between DTE and DTU, considering a set of three genes (shown in red, blue and green) consisting of 2, 2 and 3 transcripts. In the case of DTE (upper panel) the overall expression of transcripts $1,2, 6$ and $7$ change: in particular transcripts $1$ and $2$ are up-regulated in condition A while trancripts $6$ and $7$ are up-regulated in condition B. In the lower panel of Figure \ref{fig:dteVSdtu} note that only transcripts $6$ and $7$ are DTE. However, also note that now the relative expression of these transcripts conditionally on the set of the same-gene transcripts (green color) is not the same between conditions. In general, DTU implies DTE but the reverse is not necessarily true.

In this paper we extend the use of two available methods in order to perform Bayesian inference for the problem of DTU. cjBitSeq \citep{cjBitSeq} was originally introduced as a Bayesian read-based model for DTE inference and here we modify it for the DTU problem. We also propose a Bayesian version of DRIMSeq \citep{drimseq2}, a count-based approach originally introduced as a frequentist model for DTU inference. Genome-scale studies incorporate a large number of multiple tests, typically at the order of tens of thousands. A crucial issue under a multiple comparisons framework is the control of the False Discovery Rate (FDR), that is, the expected proportions of errors among the rejected hypotheses \citep{benjamini1995}. According to a recent benchmarking study \citep{Soneson025387}, the ability of frequentist count-based methods to control the FDR is drastically improved by pre-filtering low-expressed transcripts. This remains true for the Bayesian version of the count-based method presented here (DRIMSeq). However it is not possible to incorporate such a strategy for read-based methods (cjBitSeq) where transcript expression levels are not known a priori. Therefore, under our Bayesian framework, we also propose the use of transformations of the raw posterior probabilities and filtering the output based on the notion of trust regions which are motivated from realistic scenarios of gene regulation \citep{gonzalez2013transcriptome}.

The rest of the paper is organized as follows. In Section \ref{sec:previous} we briefly describe existing methods. The proposed Bayesian models are presented in Section \ref{sec:Bayesian}. More specifically, Section \ref{sec:cjBitSeq} reviews the cjBitSeq framework and also introduces the necessary prior modifications for the problem of DTU. The likelihood of the DRIMSeq model is presented in Section \ref{sec:BayesDRIMSeq} and a Bayesian version is introduced next, along with a  detailed description of the inference. Section \ref{sec:fdr} deals with FDR control procedures. In Section \ref{sec:simulation} we report our findings on synthetic data using the carefully designed simulation study of \cite{Soneson025387}. In Section \ref{sec:simBayes} we compare cjBitSeq and BayesDRIMSeq with respect to the decision rules of Section \ref{sec:fdr} using power versus achieved FDR plots. In Section \ref{sec:simBenchmark} we benchmark these methods against existing ones and we also report more performance measures, such as ROC and precision/recall curves as well as comparisons in terms of run-time and memory requirements. A real RNA-seq dataset is analysed in Section \ref{sec:real}. The manuscript concludes with a Discussion. A prior sensitivity analysis of BayesDRIMSeq as well as a comparison between alternative inputs of BayesDRIMSeq and DRIMSeq based on different quantification methods  is  provided in the Appendix.

\section{Existing methods}\label{sec:previous}

\paragraph{cuffdiff} The cufflinks/cuffdiff \citep{cuffdif,cuffdif2} pipeline estimates the expression of a set of transcripts and then performs various differential expression tests both on the transcript
 and gene level. DTU at the gene level is based on comparing the similarity of two distributions using the square root of the Jensen-Shannon divergence \citep{js1,js2}. Following \citet{Soneson025387}, we used the gene-wise FDR estimates from the {\tt cds.diff} output file of cuffdiff (version 2.2.1).

\paragraph{DEXSeq} DEXSeq \citep{dexseq} is the most popular method for inferring DTU. The genome is divided into disjoint parts of exons (counting bins) and a matrix of read counts into the counting bins is used as input. The default method for counting reads for this purpose is HTSeq \citep{Anders15012015}. Given the estimated reads from HTSeq,  a negative binomial generalized linear model is fit and DTU is inferred by testing whether the interaction term between conditions is different from zero. 

\paragraph{DRIMSeq} This recent package \citep{drimseq2} implements a dirichlet-multinomial model in order to describe the variability between replicates. A likelihood ratio test is performed in order to compare a full model with distinct parameters per condition and a null model which assumes that the parameters are shared. The input is a matrix of counts per transcript. We applied this method using the following filtering criteria:
\begin{itemize}
\item $\mbox{\tt min\_gene\_expr} = 1$ (Minimal gene expression in cpm)
\item $\mbox{\tt min\_feature\_prop} = 0.01$  (Minimal proportion for feature expression) 
\item $\mbox{\tt min\_samps\_gene\_expr} = 3$ (Minimal number of samples where genes should be expressed)
\item $\mbox{\tt min\_samps\_feature\_prop} = 3$ (Minimal number of samples where features should be expressed)
\end{itemize}

\paragraph{edgeR} The function {\tt spliceVariants} from the edgeR \citep{edger} package can be used to identify genes showing evidence of splice variation using negative binomial generalized linear models. For each gene (containing at least two transcripts) a likelihood ratio test compares a model with an interaction term between each condition against a null model with no interaction term. The input corresponds to a matrix of counts per transcript.

\paragraph{limma} The function {\tt diffSplice} from the limma \citep{Ritchie20012015} package also tests for DTU by fitting negative binomial generalized linear models and performing a likelihood ratio test at the difference of log-fold changes. The input corresponds to a matrix of counts per transcript.
\section{New Bayesian approaches}\label{sec:Bayesian}

cjBitSeq was originally applied to problem of inferring transcripts with DTE and here this model is modified for the problem of DTU. DRIMSeq is a frequentist-based approach for the problem of DTU and this model is now extended under a Bayesian framework. cjBitSeq is a read-based model, that is, the observed data is a matrix of alignments of each read to the transcriptome. On the other hand, DRIMSeq is a count-based model, which uses as input a matrix of (estimated) counts corresponding to the number of reads originating from each transcript. Both methods report an estimate of the posterior probability of DTU per gene. cjBitSeq performs collapsed Gibbs sampling on the space of latent states of each transcript, that is, a binary vector with 0 corresponding to equally expressed (EE) transcripts and 1 otherwise. Bayesian DRIMSeq estimates the Bayes factor between a DTU and a null model. Therefore, cjBitSeq also reports a posterior probability of DTU for each transcript which may be of interest for transcript-level analysis. In this study we focus our attention at the gene-level summaries as done in \citet{Soneson025387}.

Both models take advantage of distributions with richer covariance structures compared to standard sampling schemes: in particular, the Generalized Dirichlet distribution is arising as a full conditional distribution at the cjBitSeq model, while DRIMSeq is based on the Dirichlet-Multinomial distribution. The Generalized Dirichlet distribution allows for positive correlations between proportions, something that it is not the case for a standard Dirichlet model, and the Dirichlet-Multinomial distribution exhibits extra variation compared to a multinomial model. Interestingly, we note that both distributions were introduced by the same author \citep{mosimann1962compound,connor}. 

\subsection{cjBitSeq}\label{sec:cjBitSeq}

Let $\bs x = (x_1,\ldots,x_r)$, $x_i\in\mathcal X$, $i = 1,\ldots,r$, denote a sample of $r$ short reads aligned to a given set of $K$ transcripts. The sample space $\mathcal X$ consists of all sequences of letters A, C, G, T. Assuming that reads are independent, the joint probability 
density function of the data is written as 
\begin{equation}\label{standardBitseq}
\bs x|\bs\theta\sim \prod_{i=1}^{r}\sum_{k=1}^{K}\theta_kf_k(x_i).
\end{equation}
The number of components ($K$) is equal to the number of transcripts and it is considered as known since the transcriptome is given. The parameter vector $\bs\theta = (\theta_1,\ldots,\theta_K)\in\mathcal P_{K-1}$ denotes relative  abundances, where 
$$\mathcal P_{K-1}:=\{p_k\geqslant 0, k=1,\ldots,K-1:\sum_{k=1}^{K-1}p_k\leqslant 1;p_K:=1-\sum_{k=1}^{K-1}p_k\}.$$
The component specific density $f_k(\cdot)$ corresponds to the probability of a read aligning at some position of transcript $k$, $k=1,\ldots,K
$. Since we assume a known transcriptome, $\{f_k\}_{k=1}^{K}$ are known as well and they are computed according to the methodology described in
 \cite{bitseq}, taking into account optional position and sequence-specific bias correction method.

\citet{cjBitSeq} proposed a Bayesian model selection approach for identifying differentially expressed transcripts from RNA-seq data. The methods builds upon the BitSeq model \citep{bitseq,papVB,bitseqVB}. Compared to other approaches, the main difference of cjBitSeq is that transcript expression and differential expression is jointly modelled. In contrast to other methods where the starting point of the DE analysis is a count matrix, the input of cjBitSeq is the matrix $L$ containing alignment probabilities of each read to the transcriptome. According to Equation \eqref{standardBitseq}, the probability of read $i$ aligning at transcript $k$ is given by $L_{ik} = f_k(x_i)$ for $ i =1,\ldots, r$ and $k = 1,\ldots,K$.

Assume that we have at hand two samples $\bs x := (x_1,\ldots,x_{r})$ and $\bs y:=(y_1,\ldots,y_{s})$, with $r$ and $s$ denoting the number of (mapped) reads for sample $\bs x$ and $\bs y$, respectively. Now, let $\theta_k$ and $w_k$ denote the unknown relative abundance of transcript $k=1,\ldots,K$ in sample $\bs x$ and $\bs y$, respectively. Define the parameter vector of relative abundances as
$\bs\theta = (\theta_1,\ldots,\theta_{K-1};\theta_K)\in\mathcal P_{K-1}$
and $\bs w = (w_1,\ldots,w_{K-1};w_K)\in\mathcal P_{K-1}$. Under the standard BitSeq model the prior on the parameters $\bs \theta$ and $\bs w$ would be a product of independent Dirichlet distributions. In this case the probability $\theta_k= w_k$ under the prior is zero and it is not straightforward to define non-DE transcripts. To model differential expression we would instead like to identify instances where transcript expression has not changed between samples. Therefore, we introduce a finite probability for the event $\theta_k=w_k$. This leads us to define a new model with a non-independent prior for the parameters $\bs\theta$ and $\bs w$.

\begin{mydef}[State vector]\label{def:state} Let $c:=(c_1,\ldots,c_K)\in\mathcal C$, where $\mathcal C$ is the set defined by:
\begin{enumerate}
\item $c_k\in\{0,1\}$, $k=1,\ldots,K$
\item $c_+:=\sum_{k=1}^{K}c_k\neq 1$.
\end{enumerate}
Then, for $k=1,\ldots,K$ let:
$
\begin{cases}
\theta_k = w_k, & \text{if }c_k=0\\
\theta_k \neq w_k, & \text{if }c_k=1.
\end{cases}
$
\noindent
We will refer to vector $c$ as the state vector of the model.
\end{mydef}

cjBitSeq was originally applied to the problem of DTE by introducing a cluster representation of aligned reads to transcripts. This clustering approach substantially reduces the dimensionality of the samp
ling space and makes the MCMC sampler converge to reasonable time. It is important to mention that clusters are defined under a data-driven algorithm, that is, by searching the alignments of each read and
 identifying groups of transcripts sharing reads. 

Under the same approach, we would be able to infer clusters of transcripts with DTU. However, since in this work we focus on inference at the gene level, we impose the assumption that clusters  are defined as the transcripts of each gene. Otherwise, in some instances it will not be straightforward to perform inference at the gene level, due to the possibility of clusters of transcripts merging multiple genes together. For example,  we found that approximately $4.5\%$ of mapped reads align to more than one gene in our simulation experiments of Section \ref{sec:simulation} using paired-end reads with length 101 base-pairs. In case that a read maps to more than one gene, we only keep the alignments corresponding to transcripts of the gene containing the best score for this specific read. Thus, the cjBitSeq algorithm is applied separately to each gene (consisting of at least two transcripts). 

For the problem of DTU, cjBitSeq is applied under a modification in the prior distribution of DE per transcript. Under the Jeffreys' prior, which is used in the default cjBitSeq setting, the probability of a gene consisting of DE transcripts is an increasing function of the number of transcripts. This prior is reasonable at a transcript-level analysis and it has been shown that it outperforms other choices. However, this choice introduces a prior bias to the case of DTU since genes with larger number of transcripts are assigned larger prior probability of DTU than genes with small number of transcripts. Therefore, now it is a priori assumed that the probability of no differential expression within a gene is equally weighted with the event that at least two transcripts exhibit DTU, that is, $\mathbb P(c_+ = 0) = 0.5$. An equal prior probability is assigned to the rest possible configurations. Thus, the prior distribution on the state vector is defined as: 
\begin{equation}\label{eq:c_prior}
\mathrm{P}(c)=\mathrm{P}(c|c_+\neq 1)=\begin{cases}
0.5, & c_+ = 0\cr
\frac{0.5}{2^{K}-K-1}, & c_+ \geqslant 2.
\end{cases}
\end{equation}
This modification is necessary in order to ensure that no prior bias is enforced at the gene-level which is the aim of the analysis in the DTU setup. 

\begin{figure}[t]
\centering
\includegraphics[scale=0.35]{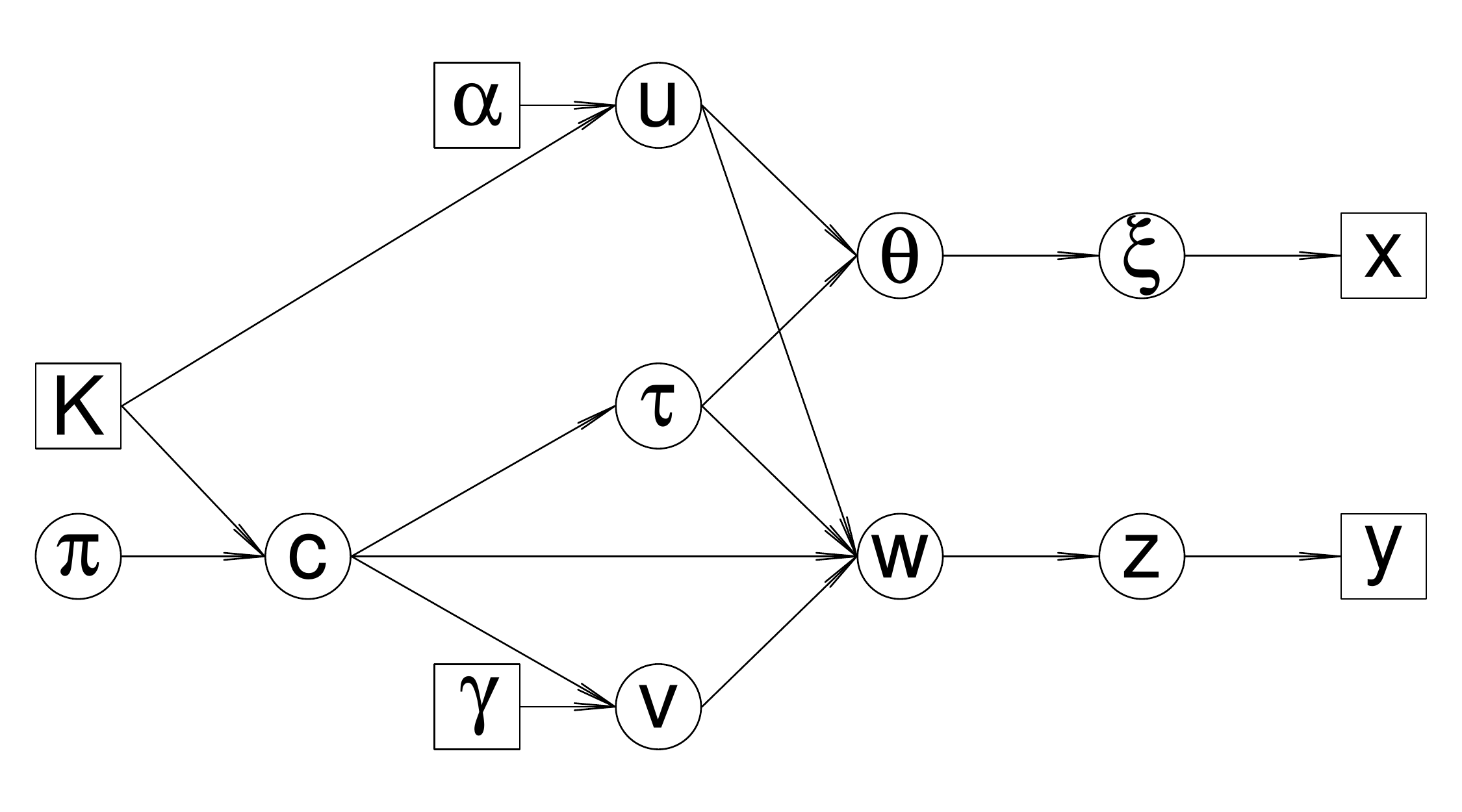}
\caption{Directed Acyclic Graph representation for the cjBitSeq model. Squares and circles represent unknown and observed/fixed quantities, respectively.}\label{fig:dag_cj}
\end{figure}

A graphical model of the cjBitSeq prior assumptions is shown in Figure \ref{fig:dag_cj}. The binary state vector $c = (c_1,\ldots,c_K)$ defines differentially or equally expressed transcripts within each gene. The prior distribution of $\bs c$ is given by Equation \eqref{eq:c_prior}, although in the general implementation of \cite{cjBitSeq} an extra level of hierarchy is imposed by the hyper-parameter $\pi$, shown in Figure \ref{fig:dag_cj}.  The parameters $\bs u$ and $\bs v$ are a-priori independent Dirichlet random variables. The dimension of $\bs u$ is equal to $K$, i.e.~the number of transcripts for a given gene. On the other hand, $\bs v$ is a random variable with varying dimension, which is defined by the number of differentially expressed transcripts, that is, $\sum_{k=1}^{K}c_k$. The parameters $\bs u$ and $\bs v$ along with an auxiliary parameter $\bs \tau$ define via a suitable one-to-one transformation the actual transcript expression parameters $\bs\theta$ and $\bs w$. According to Theorem 1 of \cite{cjBitSeq}, $\bs \theta$ and $\bs w$ are marginally Dirichlet random variables, however they are not independent since the probability of the events $\{\theta_k = w_k; k = 1,\ldots,K\}$ is positive. At the next level of hierarchy, the latent allocation variables $\bs\xi$ and $\bs z$ define the transcript allocation of each read from sample $\bs x$ and $\bs y$, respectively, through the equations $P(\xi_i = k) = \theta_k$, independent for $i = 1,\ldots,r$, and  $P(z_j = k) = w_k$, independent for $j = 1,\ldots, s$.

\citet{cjBitSeq} showed that the model is conjugate given $c$. But in order to update $(\bs c, \bs v)$, a reversible-jump mechanism \citep{Green:95,Richardson:97,papRJ} is required. However, this step can be avoided by analytical integration of $(\bs u, \bs v)$. Thus, a collapsed Gibbs sampler \citep{geman,gelfand,liu1994collapsed,liu1995covariance} updates the latent allocation variables ($\bs\xi$ and $\bs z$) of each read to its transcript of origin as well as the binary variables $c_k$ of each transcript state (DE or EE).  Let $\bs x_{-[i]}$ denote the vector arising from $x$ after excluding its $i$-th entry. A pseudo-code description of the collapsed Gibbs MCMC  sampler is:
\begin{enumerate}
\item Update allocation variables for sample $\bs x$: $\xi_i|\bs\xi_{[-i]},\bs z,c, \bs x, \bs y$, $i = 1,\ldots,r$.
\item Update allocation variables for sample $\bs y$: $z_j|\bs\xi,\bs z_{[-j]},c,\bs x, \bs y$, $j = 1,\ldots,s$.
\item Draw a random sample (without replacement) of indices $(j_1,j_2)$ from $\{1,\ldots,K\}$ and update the block of state vector $c_{j_1,j_2}|\bs c_{-[j_1,j_2]},\bs\xi,\bs z, \bs x, \bs y$.
\item Update $(\bs\theta,\bs w,\tau,\bs u,\bs v)|c,\bs \xi,\bs z, \bs x, \bs y$ (optional).
\end{enumerate} 
Note that the update 4 is optional in the sense that it is not required by any of the previous steps, however one can include it in order to also obtain MCMC samples of the transcript expression parameters $\bs\theta$ and $\bs w$. For a detailed description of the conditional distributions involved in steps 1--4 (as well as the alternative RJMCMC sampler) see \citet{cjBitSeq}.

According to our model, it is natural to call a gene as DE if at least two transcripts exhibit DTU. Hence, the posterior probability of DTU for a gene $g$ is defined as
\begin{equation}\label{eq:cjBitSeqposterior}
p_g = \mathbb P\{c_+ > 0|\bs x,\bs y\}, \quad g =1, \ldots,G,
\end{equation}
and it is estimated by the corresponding ergodic average across the MCMC run (after burn-in).

\subsection{BayesDRIMSeq}\label{sec:BayesDRIMSeq}

Let $n = n_g$ denotes the total number of reads aligning to a gene $g$ with $k$  transcripts, $g = 1,\ldots,G$. Assume that $\bs X = \bs X_g = (X_{1},\ldots,X_{k})$ is the vector of reads originating from
 each transcript, according to an underlying vector  $\bs \theta = \bs \theta_g = (\theta_{1},\ldots,\theta_{k})$ of relative abundances which is unknown. A priori, a Dirichlet prior is imposed on $\bs \theta$ and, given $\bs\theta$, the observed reads are generated according to a multinomial distribution, that is, 
\begin{eqnarray*}
\bs \theta &\sim& \mathcal D(\delta_1,\ldots,\delta_k)\\
\bs X|\bs\theta &\sim& \mbox{Multinomial}(n,\bs\theta)
\end{eqnarray*}
Integrating out $\bs \theta$, this model leads to the Dirichlet-Multinomial \citep{mosimann1962compound} distribution:
$$\mbox{P}(\bs X = \bs x) = \binom{n}{\bs x}\frac{\Gamma(\delta_+)}{\Gamma(n + \delta_+)}\prod_{j=1}^{k}\frac{\Gamma(\delta_j + x_j)}{\Gamma(\delta_j)}, $$
where the first term in the product denotes the multinomial coefficient and $\delta_+ = \sum_{k=1}^{K}\delta_k$. We will write: $\bs X|n,\bs\delta \sim \mathcal{DM}(n,\bs\delta)$. It can be shown that $$\mathbb E\bs X = n\bs\pi$$ and $$\mbox{Var}\bs X = \left\{1+\frac{n-1}{\delta_++1}\right\}n\{\mbox{diag}(\bs\pi)-\bs\pi\bs\pi'\},$$
where $\bs\pi = \{\delta_j/\delta_+;j=1,\ldots,k-1\}$ and $\mbox{diag}(\bs\pi)$ denotes a diagonal matrix with diagonal entries equal to $\pi_1,\ldots,\pi_{k-1}$. Note that as $\delta_+\rightarrow \infty$ the variance-covariance matrix of the Dirichlet-multinomial distribution reduces to $n\{\mbox{diag}(\bs\pi)-\bs\pi\bs\pi'\}$, that is, the variance-covariance matrix of the multinomial distribution. In any other case extra variation is introduced compared to standard multinomial sampling, a well known property of the Dirichlet-multinomial distribution (see e.g.~\citet{neerchal1998large}).

Consider now that a matrix of (estimated) read counts is available for two different conditions, consisting of $n_1$ and $n_2$ replicates. Given two hyper-parameter vectors $\bs \delta_1, \bs \delta_2$, let
\begin{eqnarray*}
\bs X_{i}^{(g)}|n_{1i},\bs\delta_{1} &\sim& \mathcal{DM}(n_{1i},\bs\delta_1), \quad \mbox{independent for } i = 1,\ldots,n_1\\
\bs Y_{j}^{(g)}|n_{2j},\bs\delta_{2} &\sim& \mathcal{DM}(n_{2j},\bs\delta_2), \quad\mbox{independent for } j = 1,\ldots,n_2,
\end{eqnarray*}
where $\bs X_{i}^{(g)}$, $\bs Y_{j}^{(g)}$ denote two independent vectors of (estimated) number of reads for the transcripts of gene $g = 1,\ldots,G$ for replicate $i = 1,\ldots,n_1$ and $j=1,\ldots,n_2$ for the first and second condition, respectively. Obviously, $n_{1i}$ and $n_{2j}$ denote the total number of reads generated from gene $g$ for the first and second condition for replicates $i$ and $j$.

In this context, DTU inference is based on comparing the hyper-parameters of the Dirichlet-Multinomial distribution. Note that $\bs\delta_1$ and $\bs\delta_2$ is proportional to the average expression level of the specific set of transcripts. Typically, there are large differences in the scale of these parameters, thus their direct comparison does not reveal any evidence for DTU. For this reason, it is essential to reparametrize the model as follows:
\begin{eqnarray}
\bs\delta_1 &=& d_1\bs g_1\\
\bs\delta_2 &=& d_2\bs g_2,
\end{eqnarray}
where $d_1>0$, $d_2>0$ and $\bs g_1 = (g_{11},\ldots,g_{1k})$, $\bs g_2 = (g_{21},\ldots,g_{2k})$, with $\sum_{i=1}^{k}g_{1i} = \sum_{i=1}^{k}g_{2i} = 1$ and $g_{1i}, g_{2i}>0$, $i = 1,\ldots,k$. 

In this case, DTU inference is based on comparing the null model: 
$$\mathcal M_0: \bs g_1 = \bs g_2$$
versus the full model where 
$$\mathcal M_1:\bs g_1 \neq \bs g_2.$$
A likelihood ratio test is implemented in the DRIMSeq package for testing the hypothesis of the null versus the full model. In this work, we propose to compare the two models by applying approximate Bayesian model selection techniques. In particular, a priori it is assumed that
\begin{eqnarray}\label{eq:exponential}
d_i&\sim&\mathcal E(\lambda), \quad\mbox{independent for } i = 1,2\\
\bs g_i &\sim& \mathcal D(1,\ldots,1)\quad\mbox{independent for } i = 1,2,
\end{eqnarray}
and furthermore $d_i$ and $\bs g_j$ are mutually independent.

In order to perform Bayesian model selection, the Bayes factor \citep{kassRaftery} of the null against the full model is approximated using a two stage procedure. At first, the posterior distribution of each model is approximated using Laplace's approximation \citep{laplace1774,laplace1986}, a well established practice for approximating posterior moments and posterior distributions \citep{tierney1986accurate,tierney89,azevedo1994laplace,raftery1996approximate}. Then, the logarithm of marginal likelihoods of $\mathcal M_0$ and $\mathcal M_1$ are estimated using independent samples from the posterior distribution via self-normalized sampling importance resampling \citep{gordon1993novel}. Finally, the posterior probabilities $p(\mathcal M_0|\bs x^{(g)},\bs y^{(g)})$, and $p(\mathcal M_1|\bs x^{(g)},\bs y^{(g)})$ are estimated assuming equally weighted prior probabilities. 

Denote by $\bs g_0$ the common value of $\bs g_1$, $\bs g_2$ in model $\mathcal M_0$. Let $\bs u_0 = (\bs g_0,d_1,d_2)\in\mathcal U_0 $, $\bs u_1 = (\bs g_1,\bs g_2,d_1,d_2)\in\mathcal U_1$ denote the parameters associated with models $\mathcal M_0$ and $\mathcal M_1$, respectively. Obviously, the underlying parameter spaces are defined as $\mathcal U_0 =\mathcal P_{K_g-1}\times(0,+\infty)^2$ and $\mathcal U_1 = \mathcal P_{K_g-1}^{2}\times(0,+\infty)^2$. The marginal likelihood of data under model $\mathcal M_j$, is defined as
\[
f(\bs x^{(g)},\bs y^{(g)}|\mathcal M_j) = \int_{\mathcal U_j}  f(\bs x^{(g)},\bs y^{(g)}|\bs u_j)f(\bs u_j|\lambda)\mathrm{d} \bs u_j,\quad j=0,1. 
\]
According to the basic importance sampling identity, the marginal likelihood model  can be evaluated using another density $\phi$, which is absolutely continuous on $\mathcal U_j$, as follows
\[
f(\bs x,\bs y|\mathcal M_j) = \int_{\mathcal U_j}  \frac{f(\bs x^{(g)},\bs y^{(g)}|\bs u_j)f(\bs u_j|\lambda)}{\phi(\bs u_j)}\phi(\bs u_j)\mathrm{d} \bs u_j.
\]
The minimum requirement for $\phi$ is to satisfy $\phi(\bs u_j)>0$ whenever $f(\bs x^{(g)},\bs y^{(g)}|\bs u_j)f(\bs u_j|\lambda) > 0$. Assume that a sample $\{\bs u^{(i)};i=1,\ldots,n\}$ is drawn from $\phi(\cdot)$. Then, the importance sampling estimate of the marginal likelihood is 
\[
\widehat{f}(\bs x^{(g)},\bs y^{(g)}|\mathcal M_j) = \frac{1}{n}\sum_{i=1}^{n}\frac{f(\bs x^{(g)},\bs y^{(g)}|\bs u_j^{(i)})f(\bs u_j^{(i)}|\lambda)}{\phi(\bs u_j^{(i)})},\quad j=0,1.
\]
The candidate distribution $\phi$ is the approximation of the posterior distribution according to the Laplace's method. It is well known that basic importance sampling performs reasonably well in cases that the number of parameters is not too large. However, it can be drastically improved using sequential Monte Carlo methods, such as sampling importance resampling \citep{gordon1993novel,liu1998sequential}. The {\tt R} package {\tt LaplacesDemon} \citep{laplaceDemon1} is used for this purpose.

Finally, the posterior probability of the DTU model is defined as
\begin{equation}\label{eq:posteriorDRIMSeq}
p_g = \mathbb P(\mathcal M_1|\bs x^{(g)},\bs y^{(g)}) \propto f(\bs x^{(g)},\bs y^{(g)}|\mathcal M_1)P(\mathcal M_1), \quad g = 1,\ldots,G,
\end{equation}
by also assuming equally weighted prior probabilities, that is, $P(\mathcal M_1) = P(\mathcal M_0) = 0.5$. Note that the Bayes Factor of the null against the full model is then given by
\begin{equation*}
B_{01}^{(g)} = \frac{\mathbb P(\mathcal M_0|\bs x^{(g)},\bs y^{(g)})}{\mathbb P(\mathcal M_1|\bs x^{(g)},\bs y^{(g)})}=\frac{f(\bs x^{(g)}, \bs y^{(g)}|\mathcal M_0)}{f(\bs x^{(g)}, \bs y^{(g)}|\mathcal M_1)},\quad g = 1,\ldots,G
\end{equation*}
since the prior odds ratio is equal to one.

In case that low expressed transcripts are included in the computation, the Laplace approximation faces many convergence problems. We have found that this problem can be alleviated by pre-filtering low expressed transcripts, as also pointed out by \cite{Soneson025387}. 

\section{Bayesian FDR control for the problem of DTU}\label{sec:fdr}
In this section we consider various decision rules in order to control the False Discovery Rate (FDR) \citep{benjamini1995,storey2003,fdrJASA,fdrISBA}. Decision rules \eqref{eq:fdrGeneProbDE} and \eqref{eq:fdrRaw} are taking into account the whole set of genes and make use of the raw and transformed posterior probabilities, respectively. Intuitively, the transformation of  posterior probabilities prioritizes genes consisting of transcripts with large changes in their expression. Decision rules \eqref{eq:fdrSwitch} and \eqref{eq:fdr} are based on filtering the output of \eqref{eq:fdrGeneProbDE} and \eqref{eq:fdrRaw} according to a trust region.

A decision rule based on the raw gene-level posterior probabilities of DTU, as defined in Equations \eqref{eq:cjBitSeqposterior} and \eqref{eq:posteriorDRIMSeq}, is the following.
\begin{equation}\label{eq:fdrGeneProbDE}
d_{1g} = \begin{cases}
1, & \widehat p_g \geqslant 1 - \alpha\\
0, & \mbox{otherwise.}
\end{cases}
\end{equation}
Note that for the problem of inferring DTE the decision rule \eqref{eq:fdrGeneProbDE} is the one used by \citet{ebseq}. However, the cjBitSeq model takes into account changes to any subset of transcripts within a gene, thus, \eqref{eq:fdrGeneProbDE} may identify a large number of genes consisting of relatively small changes in low expressed transcripts. A more conservative choice will focus our attention to the dominant transcripts, where more reads are  available and potentially the results will be more robust. 

Next we define a filtering of the output based on a ``trust region''. Let  $i$ and $j$ denote the estimated dominant transcripts in condition A and B, respectively. The trust region corresponds to the subset of genes where the relative ordering of estimated expression levels of dominant transcript switches, that is,
\[
G_0 = \{g = 1,\ldots,G: (\widehat{\theta}_i^{(g)}-\widehat{\theta}_j^{(g)})(\widehat{w}_i^{(g)}-\widehat{w}_j^{(g)})<0\}.
\]
Switching events between dominant transcripts have been proposed as a major source of DTU in real RNA-seq data \citep{gonzalez2013transcriptome}. 

Note that in the previous expression we used the notation of transcript expression levels according to cjBitSeq. For BayesDRIMSeq $\bs\theta$ and $\bs w$ should be replaced by $\bs g_1$ and $\bs g_2$, respectively. The decision rule which corresponds to filtering \eqref{eq:fdrGeneProbDE} according to $G_0$ is the following:
\begin{equation}\label{eq:fdrSwitch}
d_{2g} = \begin{cases}
1, & \widehat{p}_g \geqslant 1 - \alpha \mbox{ and } g\in G_0\\
0, & \mbox{otherwise.}
\end{cases}
\end{equation}
Note that decision rules $d_1$ and $d_2$ are solely based on the posterior probabilities of gene DTU and the trust region, respectively. However, it makes sense to also take into account additional information, such as the magnitude of the change of the within gene relative transcript expression, which is a by-product of our algorithm. 

In order to clarify this, consider the following example. Assume that genes $g_1$ and $g_2$ both consist of two transcripts. For $g_1$, let $\theta^{(g_1)}_1 = 0.1$, $\theta^{(g_1)}_2 = 0.9$ and $w^{(g_1)}_1 = 0.9$, $w^{(g_1)}_2 = 0.1$. For $g_2$, let $\theta^{(g_2)}_1 = 0.4$, $\theta^{(g_2)}_2 = 0.6$ and $w^{(g_2)}_1 = 0.6$, $w^{(g_2)}_2 = 0.4$. Furthermore, assume that the posterior evidence of DE is the same for both genes, that is, $\widehat{p}_{g_1} = \widehat{p}_{g_2} = p$.  In the case that the posterior probability $p$ is sufficiently large, genes $g_1$ and $g_2$ will be given the same importance in our discovery list. Note however that for gene $g_1$ the absolute change in relative expression is 4 times larger than for gene $g_2$. Ideally, we would like our discovery list to rank higher gene $g_1$ than gene $g_2$. This is achieved using the following FDR control procedure.

Consider any (Bayesian) method that for each gene yields an estimate of the posterior probability of DTU per gene $p_g$, $g = 1,\ldots,G$.
\begin{itemize}
\item For a given permutation $\bs\tau=(\tau_1,\tau_2,\ldots,\tau_G)$ of $\{1,2,\ldots,G\}$ and let $q_g = p_{\tau_g}$, $g=1,\ldots,G$. 
\item Define: $r_g = \frac{\sum_{j=1}^{g}1-q_j}{g}$, $g =1,\ldots,G$.
\item For $0 < \alpha < 1$, consider the decision rule:
\begin{equation}\label{eq:fdrRaw}
d_{3g} = \begin{cases}
1, & 1\leqslant g \leqslant g^{*}\\
0, & g^{*}+1\leqslant g \leqslant G
\end{cases}
\end{equation}
where $g^{*}:=\max\{g = 1,\ldots,G: r_g \leqslant \alpha\}$.
\item $\widehat{\mathbb E}(\mbox{FDR}|\mbox{data}) = \frac{\sum_{j=1}^{g^{*}}1-q_j}{g^{*}}\leqslant \alpha$
\end{itemize}

Here we mention that in the original implementation of cjBitSeq for the DTE problem, the permutation $\tau$ was defined as the one that orders the posterior probabilities of transcript DE in decreasing order.

The permutation that takes into account the previously described concept of magnitude change is defined as follows. Let $\rho_g=\max{|\widehat{\theta}_k^{(g)}-\widehat{w}_k^{(g)}|, k=1,\ldots,K_g}$, where $\widehat{\theta}_k^{(g)}$ and $\widehat{w}_k^{(g)}$ denote the posterior mean estimates of within gene transcript expression for a given transcript $k$ of gene $g$. Consider the permutation $\bs\tau = (\tau_1,\tau_2\ldots,\tau_G)$ that orders the set $\{\rho_g;g = 1,\ldots,G\}$ in decreasing order, that is:
$$\rho_{\tau_1}\geqslant \rho_{\tau_2}\geqslant \ldots\geqslant\rho_{\tau_G}.$$

Finally, we combine decision rule $d_3$ with the trust region $G_0$ to obtain our final decision rule, that is,

\begin{equation}\label{eq:fdr}
d_{4g} = \begin{cases}
1, & 1\leqslant g \leqslant g^{*} \mbox{ and }  g\in G_0\\
0, & \mbox{otherwise.}
\end{cases}
\end{equation}

\section{Simulation study}\label{sec:simulation}

\begin{figure}[t]
\centering
\begin{tabular}{c}
\includegraphics[scale=0.32]{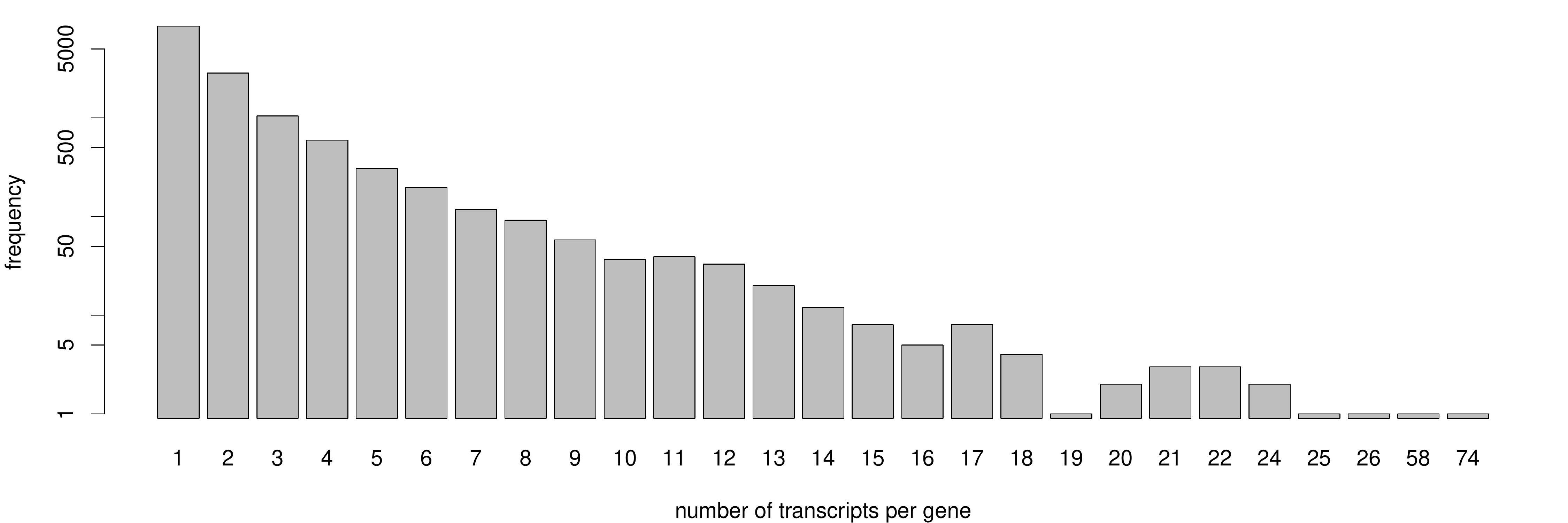}\\
\includegraphics[scale=0.32]{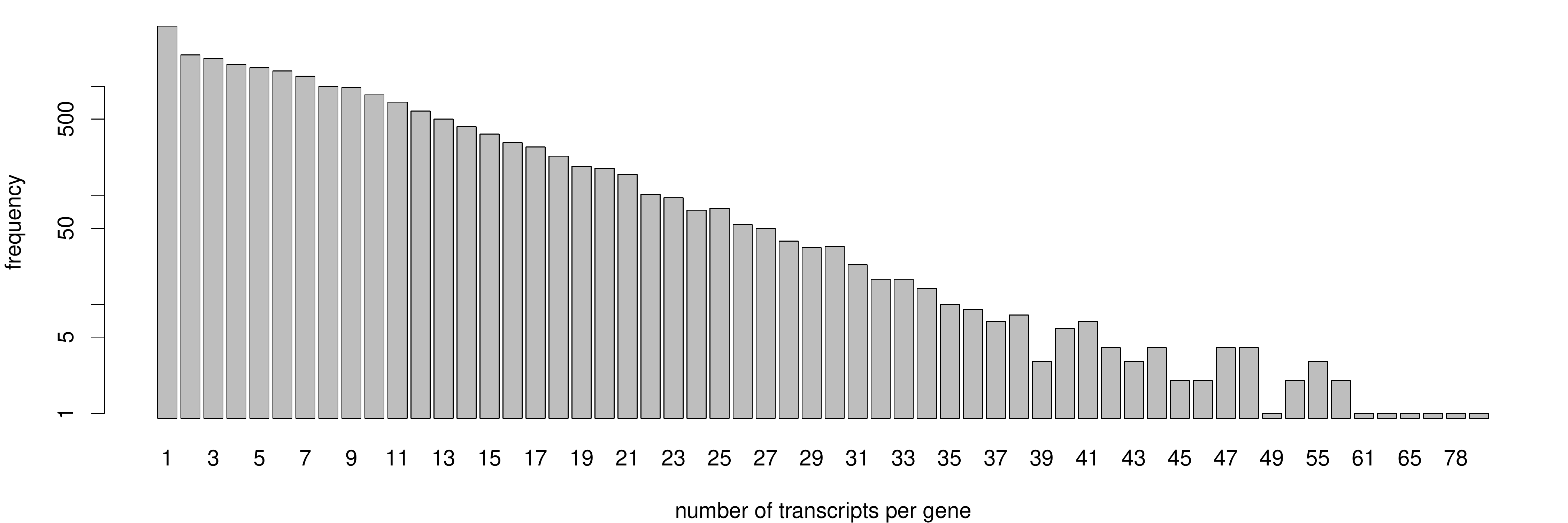}
\end{tabular}
\caption{Frequencies (in log scale) of number of annotated transcripts per gene for drosophila (up) and human (down). The total number of genes and transcripts is $13937$ and $26951$ for drosophila and $20410$ and $145342$ for human, respectively.}
\label{fig:nTr}
\end{figure}

In order to assess the performance of the proposed methods and decision rules as well as to compare against existing models, a set of simulation studies is used. Instead of setting up our own simulation scenarios, we followed the pipeline introduced in the recent study of \citet{Soneson025387}, where a large number of count-based method is being benchmarked. Synthetic RNA-seq reads are generated from the Drosophila Melanogaster and Homo Sapiens transcriptomes using the RSEM-simulator \citep{rsem}. The model parameters for RSEM-simulator were estimated from real datasets using a Negative Binomial model described in \citet{soneson2013comparison}. The transcriptomes of these two organisms exhibit strong differences as illustrated in Figure \ref{fig:nTr}. The average number of transcripts per gene is considerably smaller for fruit fly, however the transcripts are longer than for human (see also Supplementary Table 1 of \citet{Soneson025387}).

Following \citet{Soneson025387}, for each organism we simulated 3 replicates per condition. Each replicate consists of 25 million paired-end reads with length 101 base-pairs. Differential transcript usage was introduced for 1000 genes, by reversing the relative abundance of the two most abundant transcripts in one of the two conditions. The total number of reads for each transcript may or may not be equal across conditions. If the total number of reads generated from a gene is constant, no gene-level differential expression is evident. For the drosophila reads no gene-level differential expression was introduced. For human reads both cases are considered. Finally, the simulated reads are mapped to the genome or transcriptome with Tophat2 \citep{tophat} and Bowtie2 \citep{bowtie}, respectively. Cufflinks and HTSeq used the alignment files produced by Tophat2, while BitSeqVB and cjBitSeq use the alignment produced from Bowtie2, allowing a maximum of 100 hits per read. The count matrix used as input to DEXSeq is estimated using the default HTSeq method, while BitSeqVB is used for input to edgeR, limma, DRIMSeq and BayesDRIMSeq. 

\subsection{Comparison of Bayesian decision rules}\label{sec:simBayes}

\begin{figure}[p]
\centering
\begin{tabular}{ccc}
\includegraphics[scale=0.30]{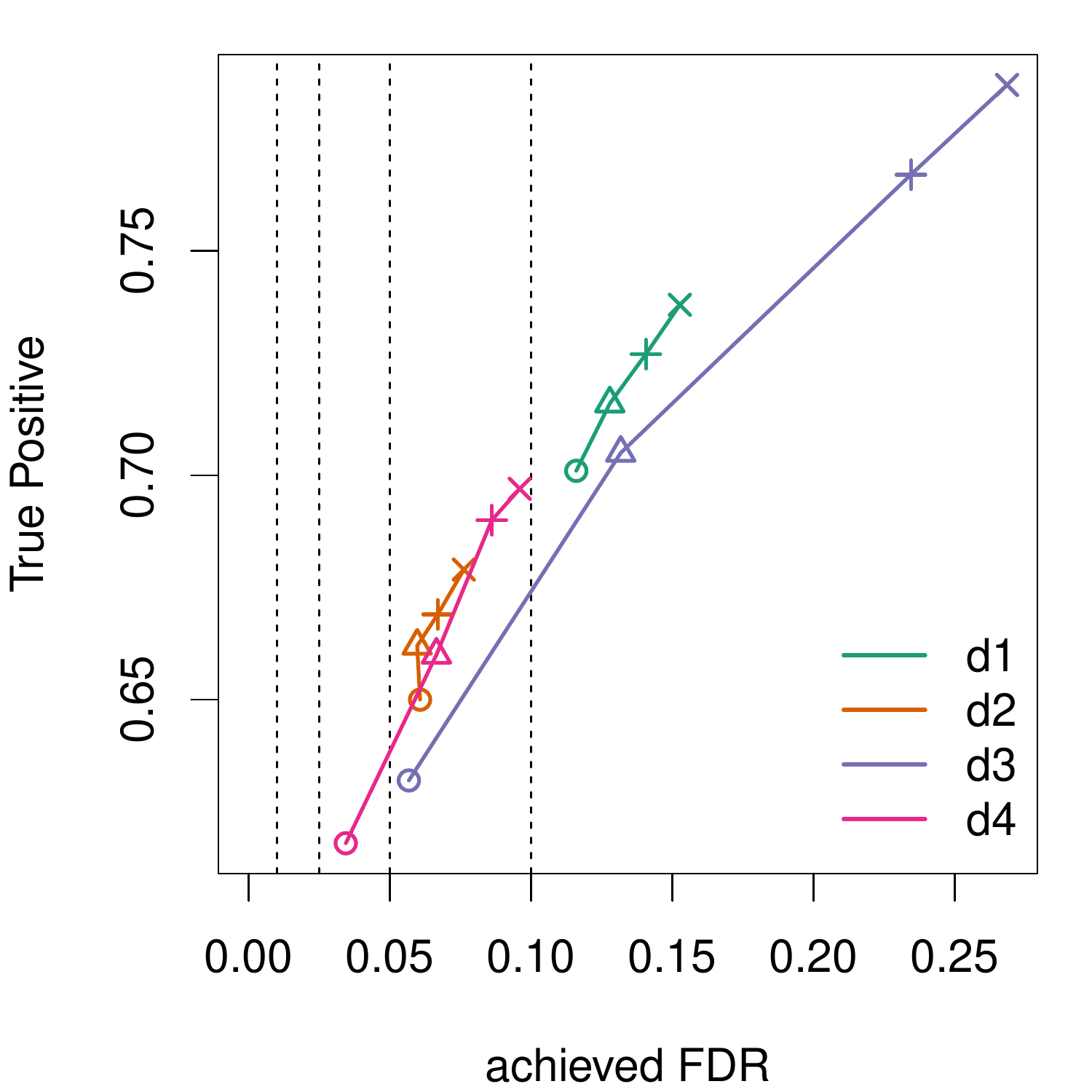}&
\includegraphics[scale=0.30]{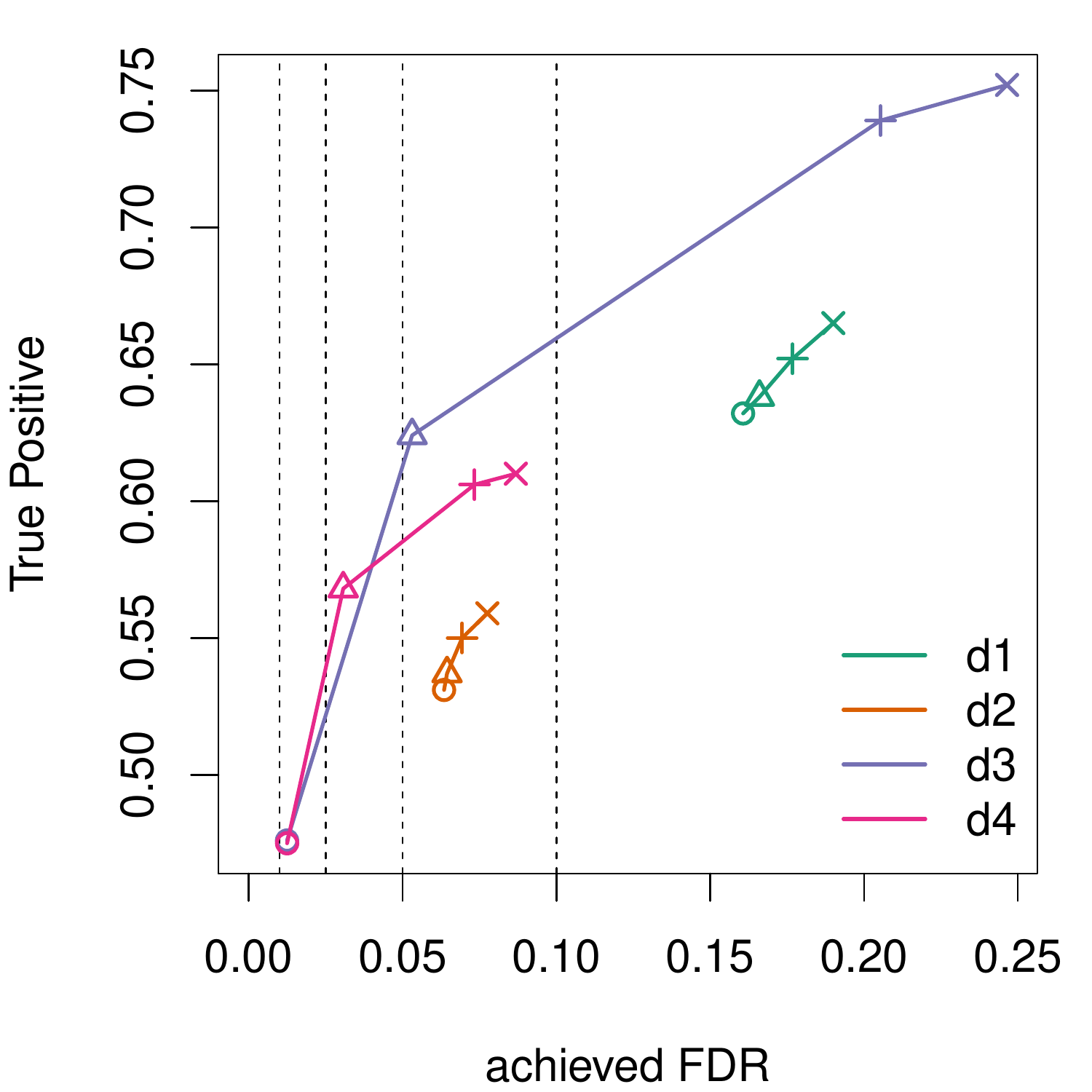}&
\includegraphics[scale=0.30]{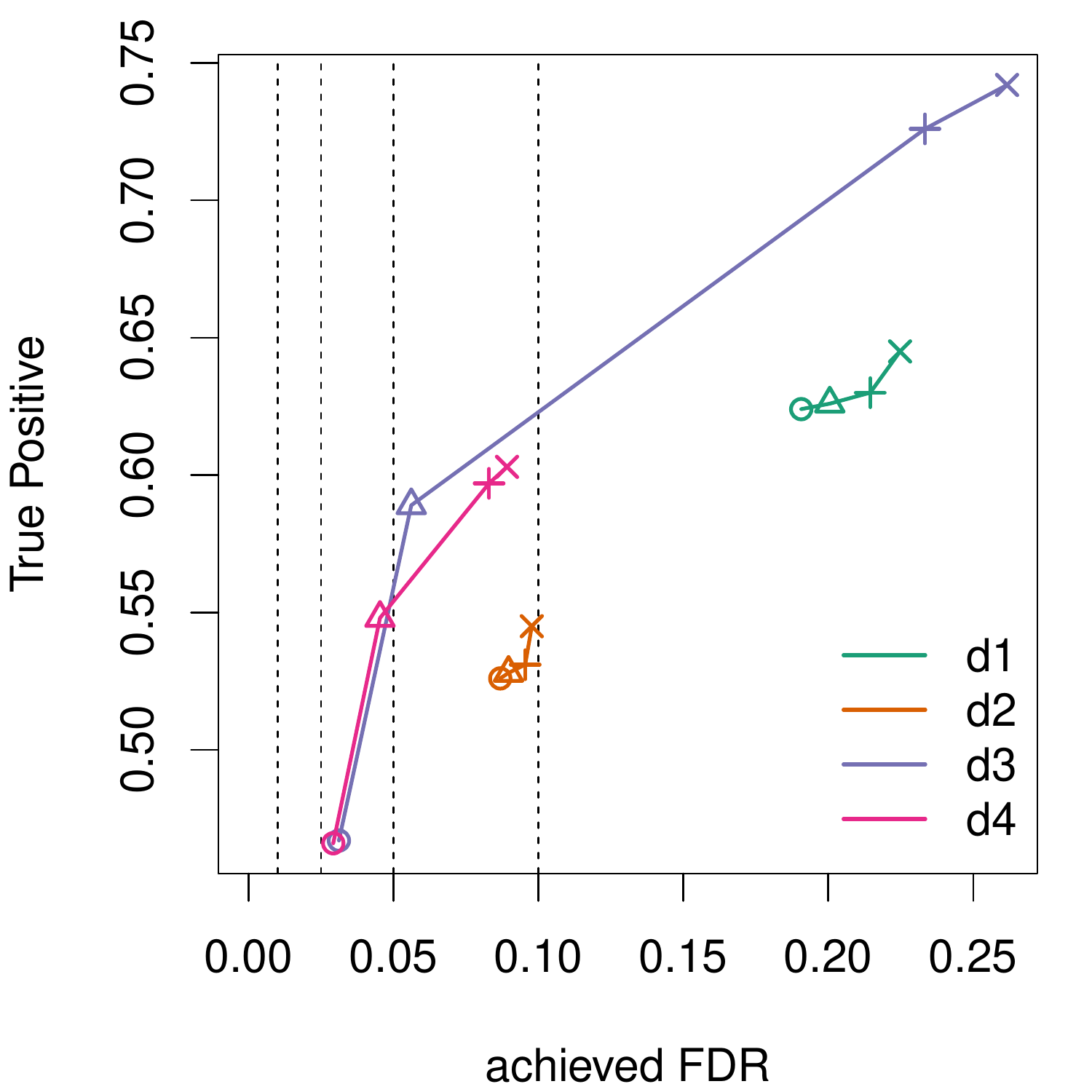}\\
\includegraphics[scale=0.30]{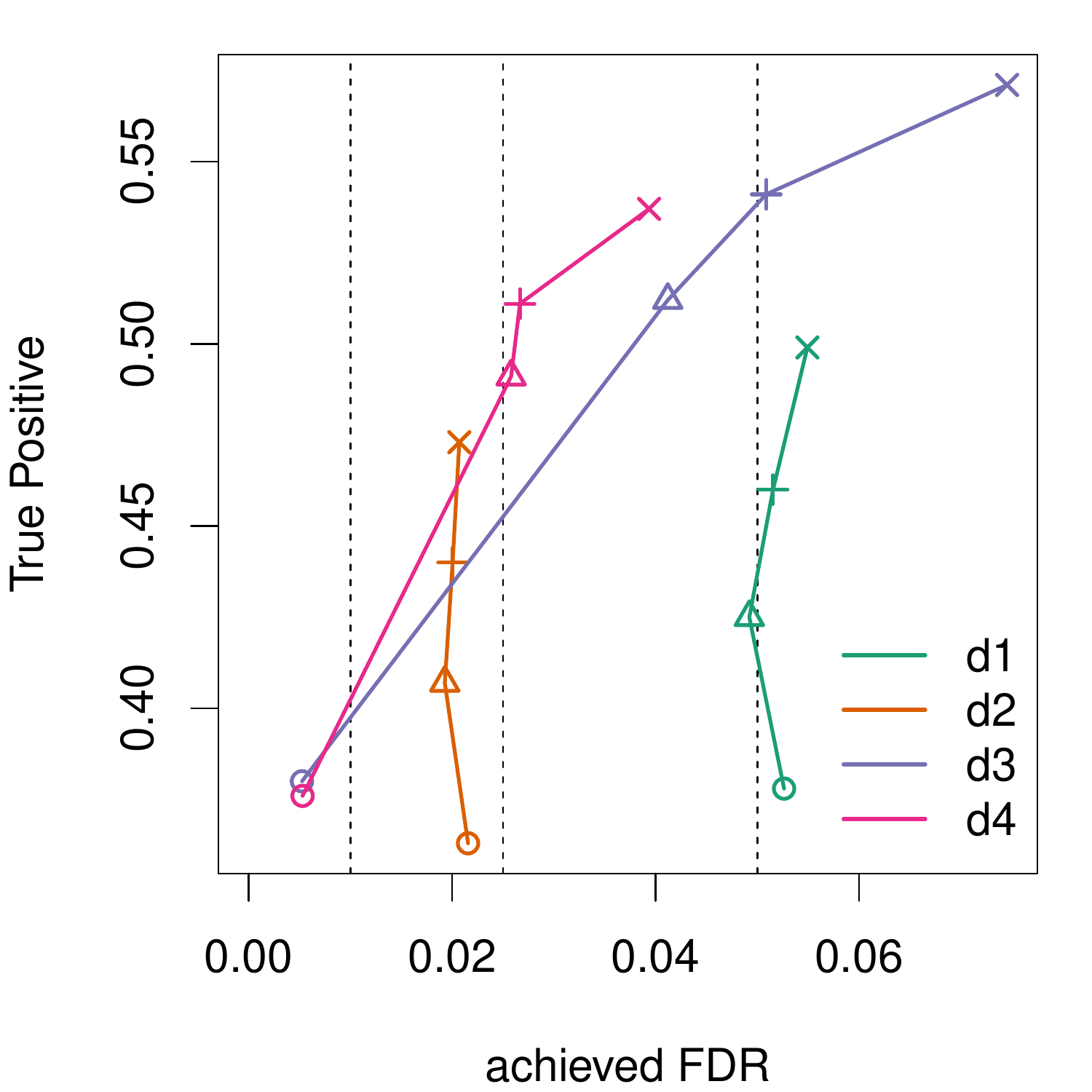}&
\includegraphics[scale=0.30]{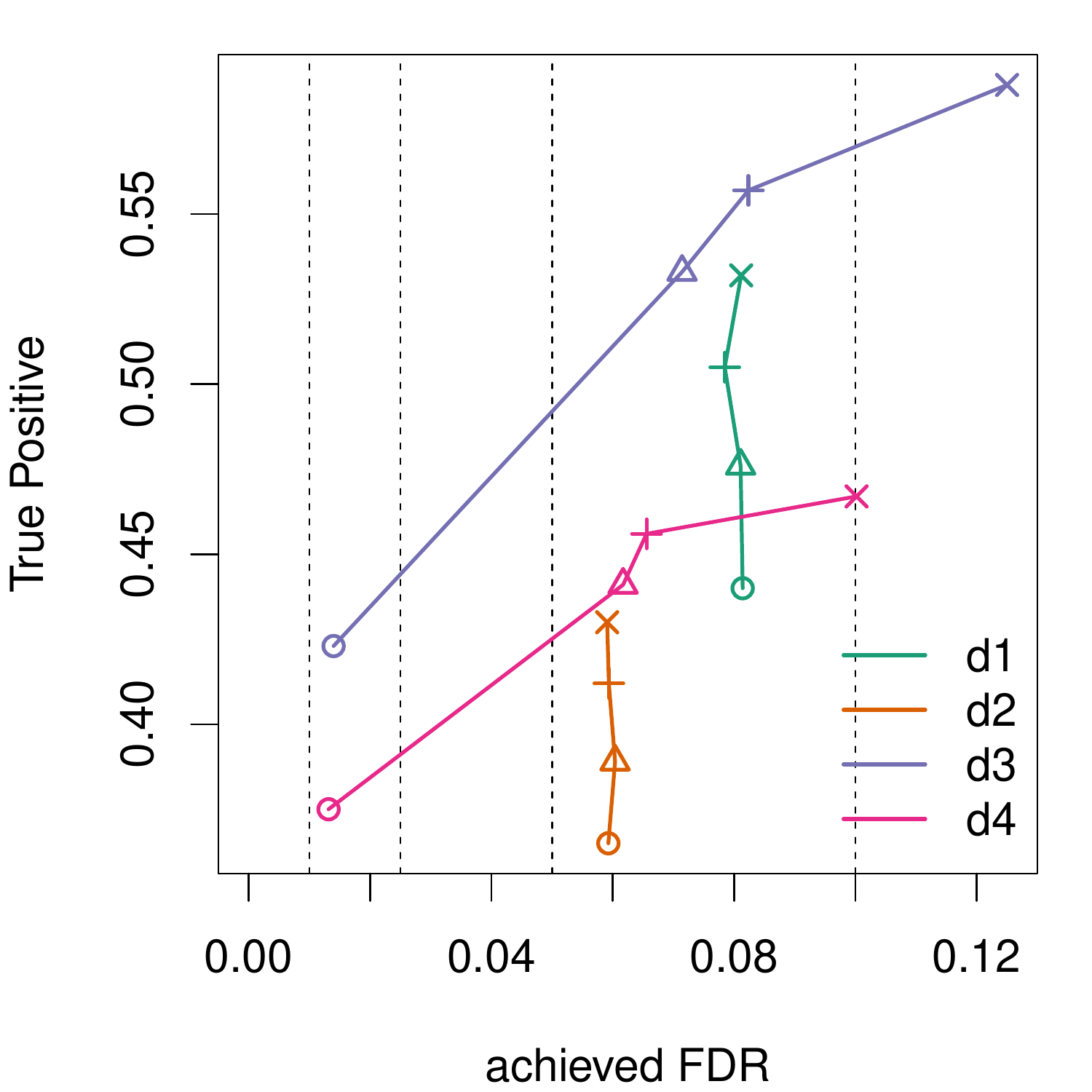}&
\includegraphics[scale=0.30]{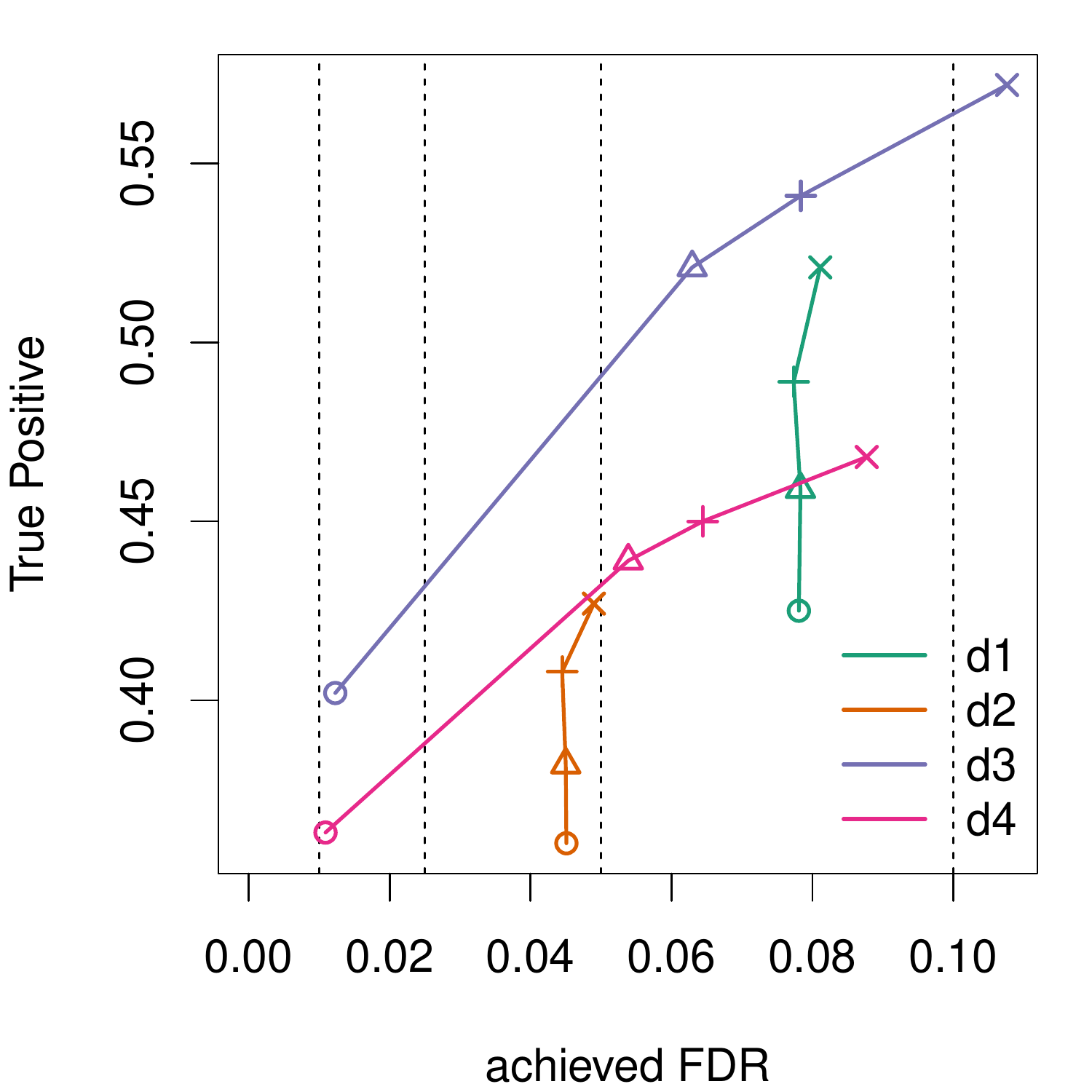}\\
\includegraphics[scale=0.30]{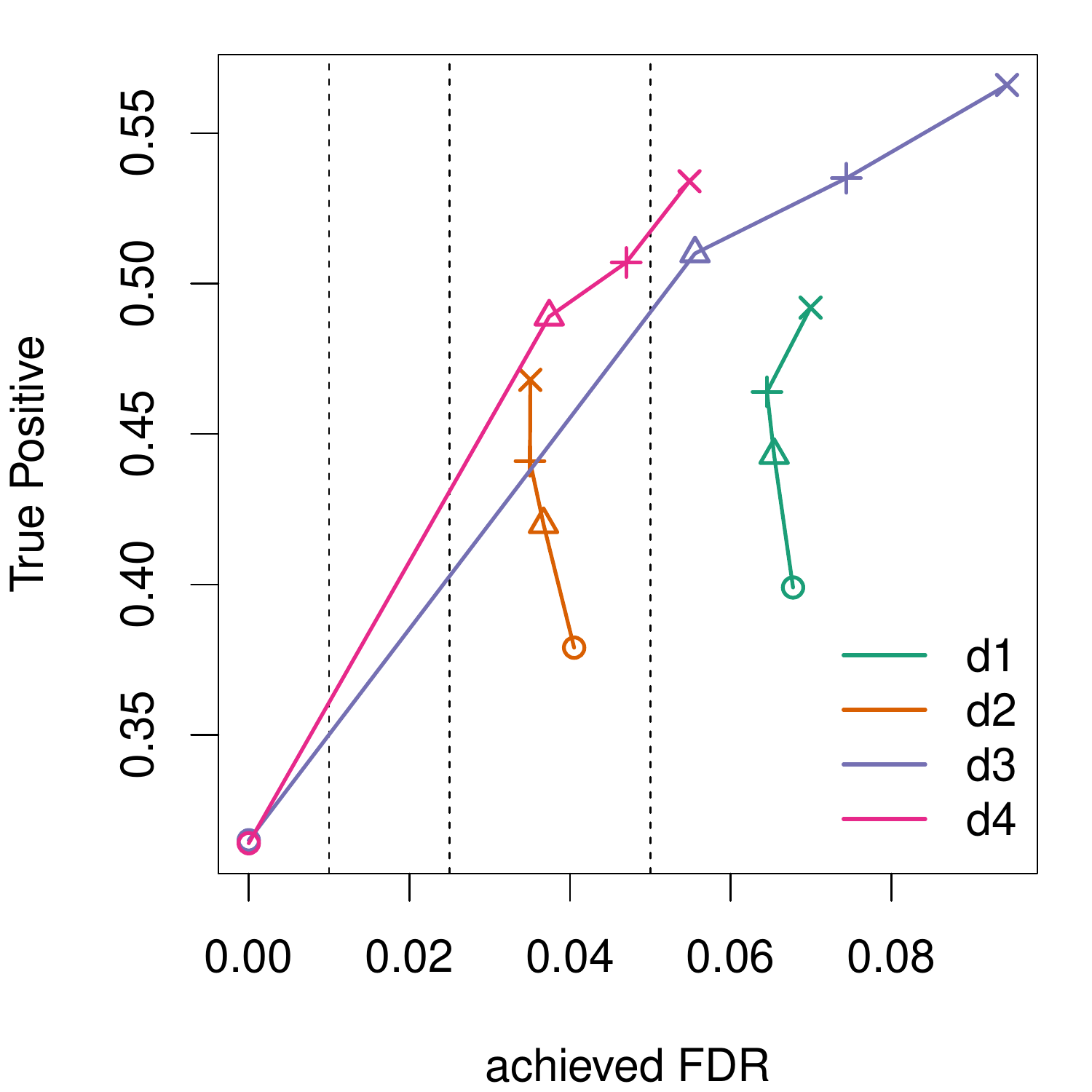}&
\includegraphics[scale=0.30]{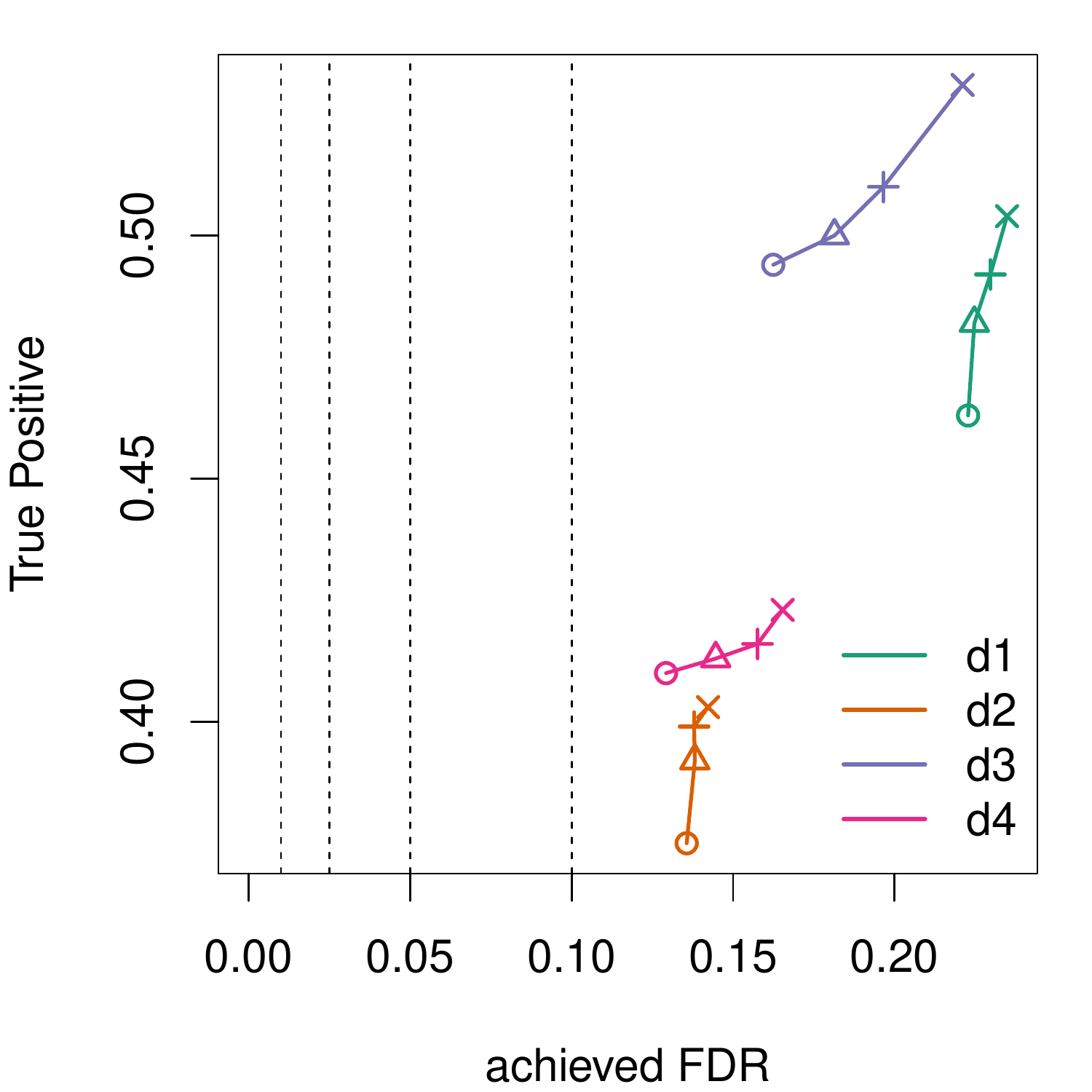}&
\includegraphics[scale=0.30]{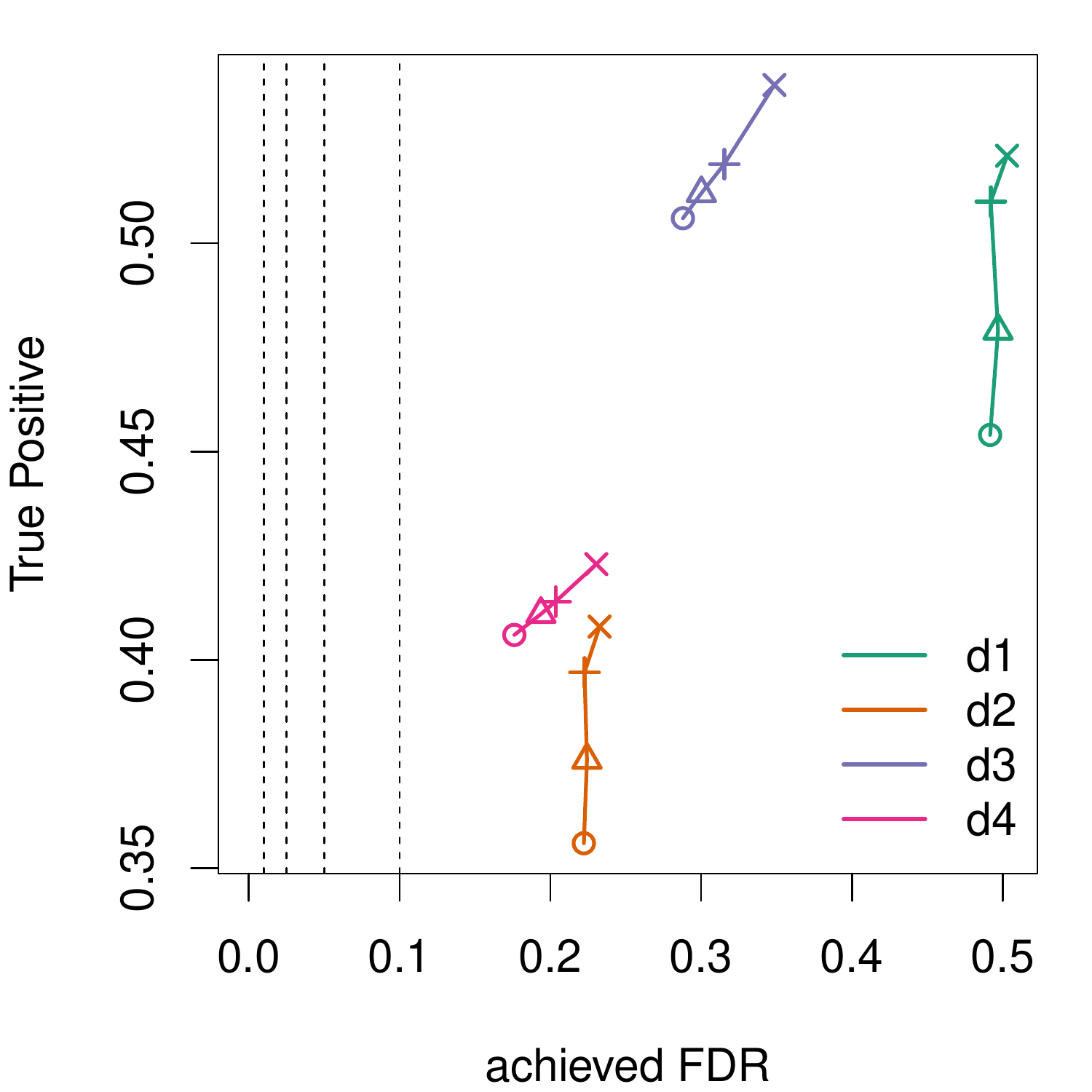}\\
(a) & (b) & (c)
\end{tabular}
\caption{Power versus achieved FDR plot using the decision rules $d_1$, $d_2$, $d_3$ and $d_4$ for cjBitSeq (1st row) and BayesDRIMSeq with (second row) and without (third row) isoform pre-filtering on the simulated data. The vertical dashed lines show the expected FDR level (0.01, 0.025,0.05,0.1). (a): drosophila, (b): human without DTE and (c): human with DTE.}
\label{fig:cjFDRs}
\end{figure}

Figure \ref{fig:cjFDRs} displays the power versus achieved FDR using the decision rules $d_k$; $k = 1,2,3,4$ for the three simulated datasets. Each rule was evaluated at four typical values of expected FDR levels, $\alpha = 0.01,0.025,0.05,0.1$, which are shown as dashed vertical lines. The plotted points correspond to the achieved FDR ($x$ axis) and the proportion of true discoveries ($y$ axis). The ability of each decision rule to control the FDR depends on the distance of each point from the corresponding vertical line: the closer, the better. On the other hand, a decision rule with higher $y$ values is more powerful. 

For cjBitSeq (upper panel) we conclude that the trust-region adjusted rules $d_2$ and $d_4$ achieve lower FDRs which are quite close to the expected values. However, note that $d_4$ yields better power compared to $d_2$, especially for the human datasets. BayesDRIMSeq is shown in middle and lower panel of \ref{fig:cjFDRs}. At the second panel of Figure \ref{fig:cjFDRs} we have applied BayesDRIMSeq by filtering out transcripts with average number of reads less than 20. The results corresponding to the full set of transcripts (no pre-filtering) are shown at the lower panel of \ref{fig:cjFDRs}. We conclude that isoform pre-filtering is essential in order to achieve reasonable control of FDR in the case of human data. Note also that under isoform pre-filtering the trust region does not have a high impact on BayesDRIMSeq.

\subsection{Comparison against existing methods}\label{sec:simBenchmark}

\begin{figure}[p]
\centering
\begin{tabular}{ccc}
\includegraphics[scale=0.30]{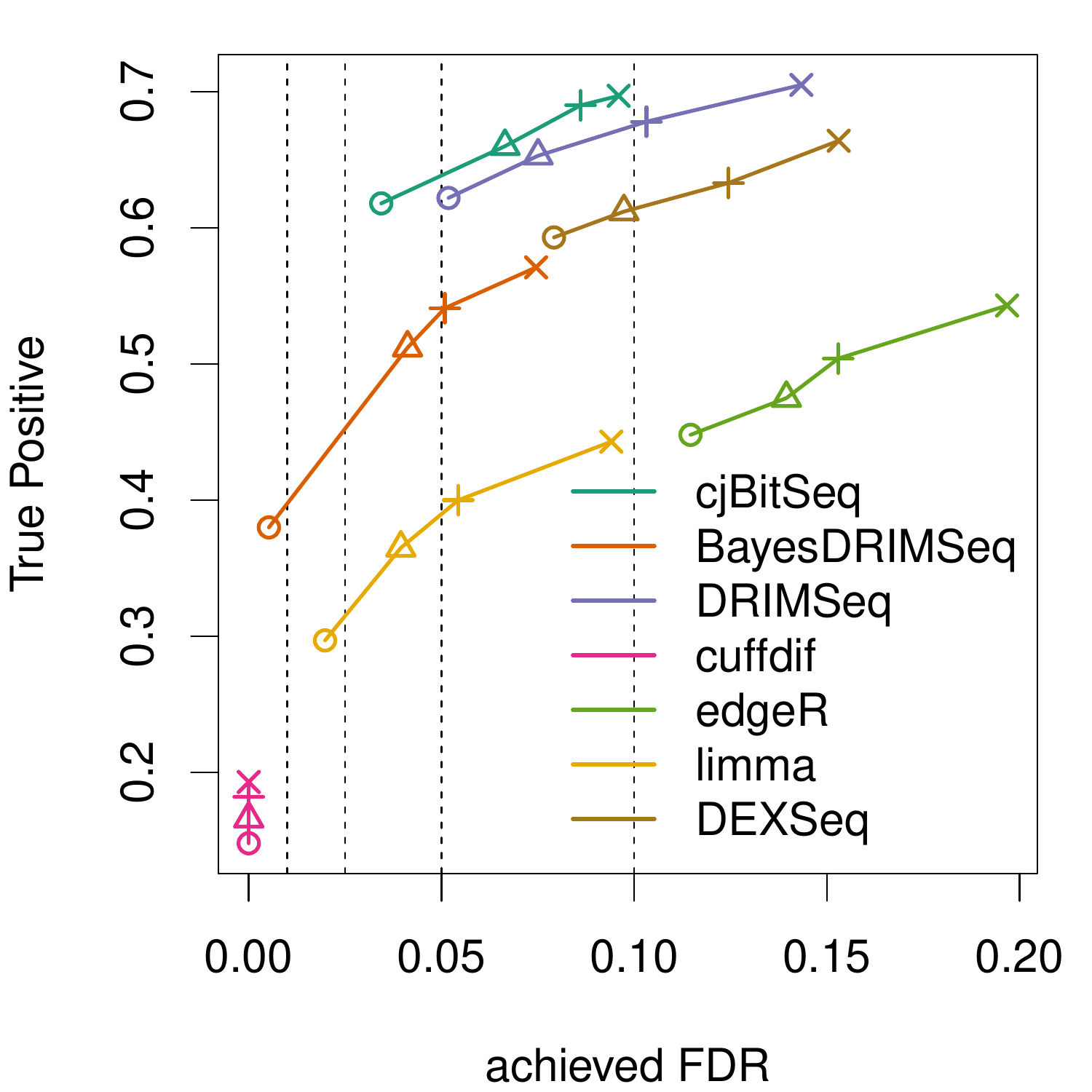}&
\includegraphics[scale=0.30]{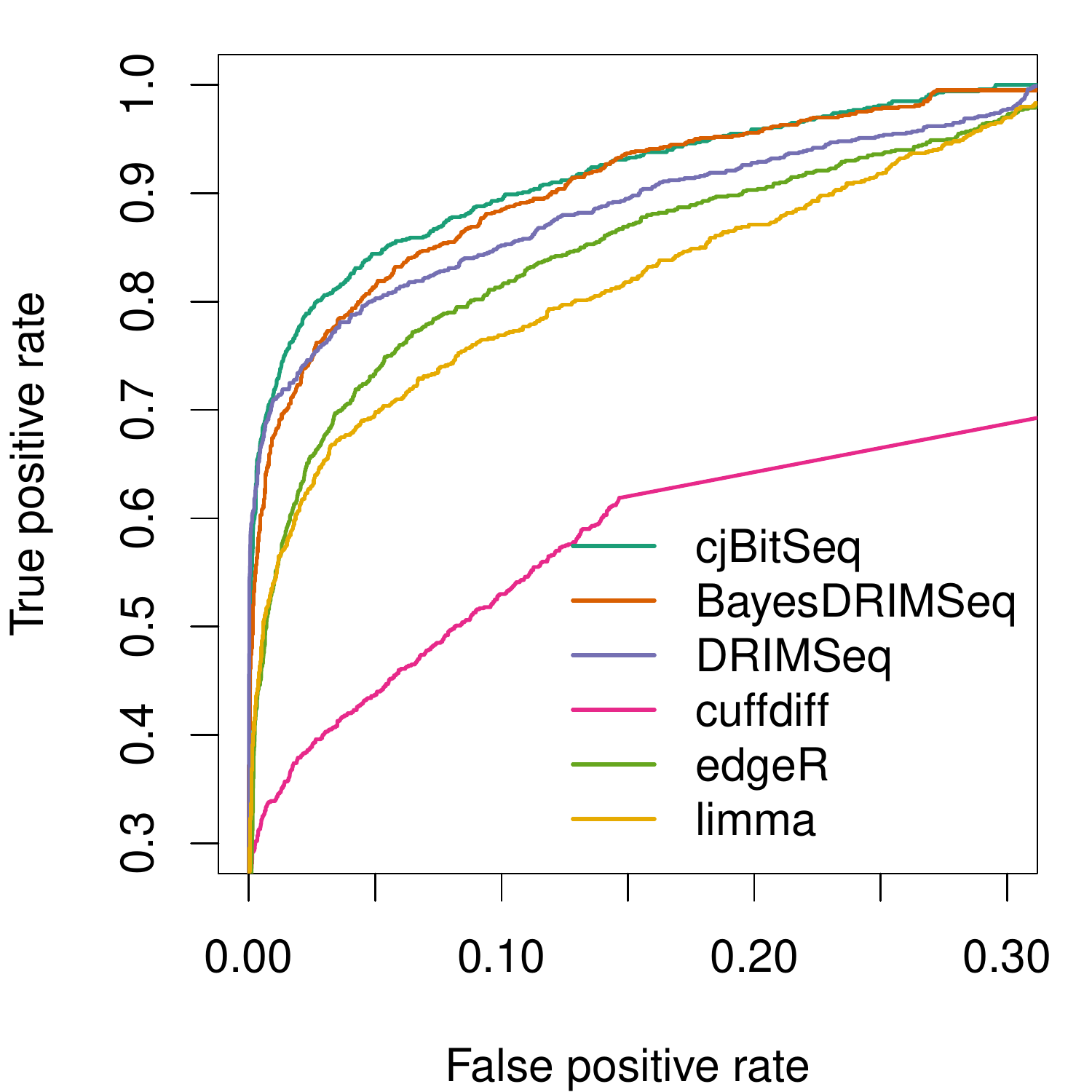}&
\includegraphics[scale=0.30]{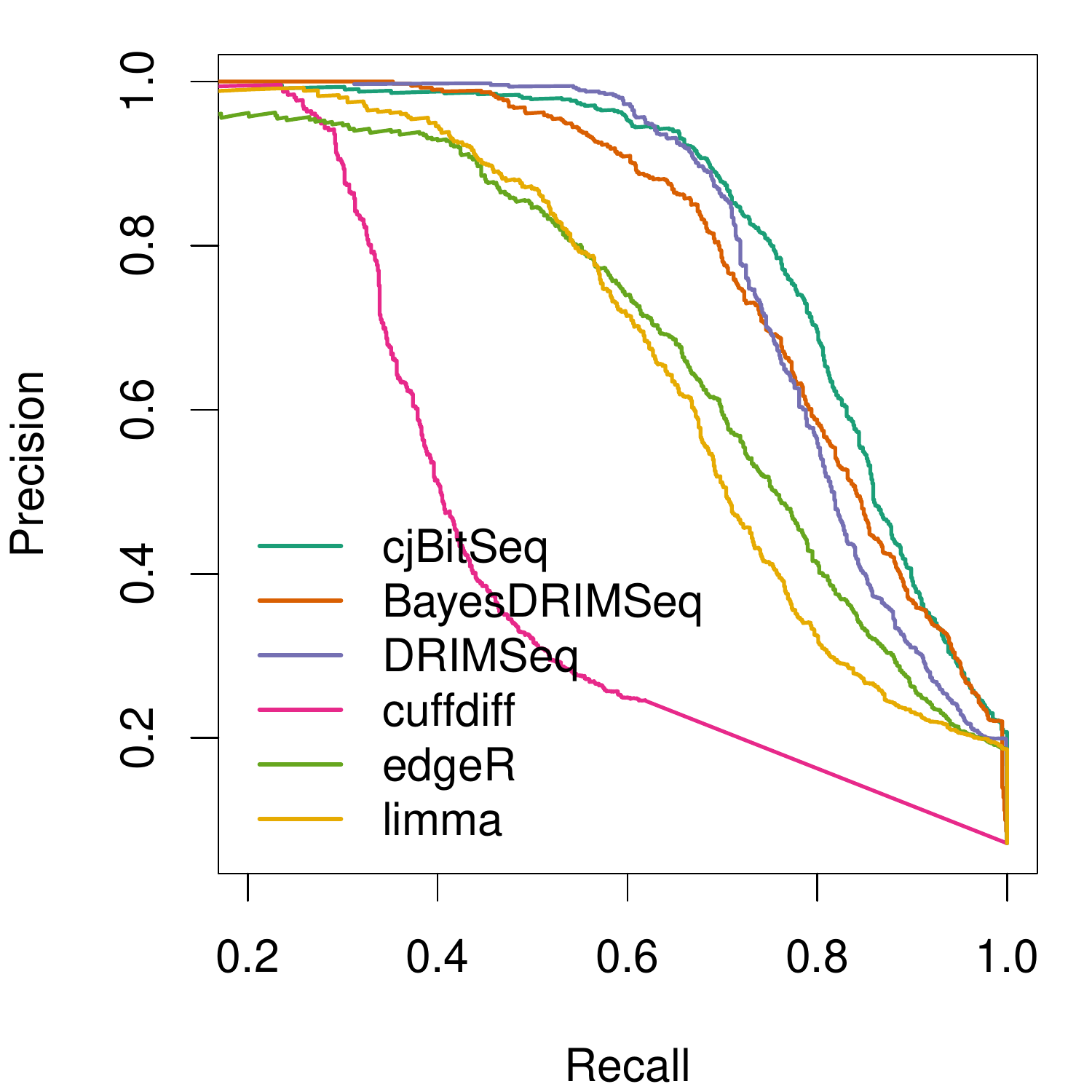}\\
\includegraphics[scale=0.30]{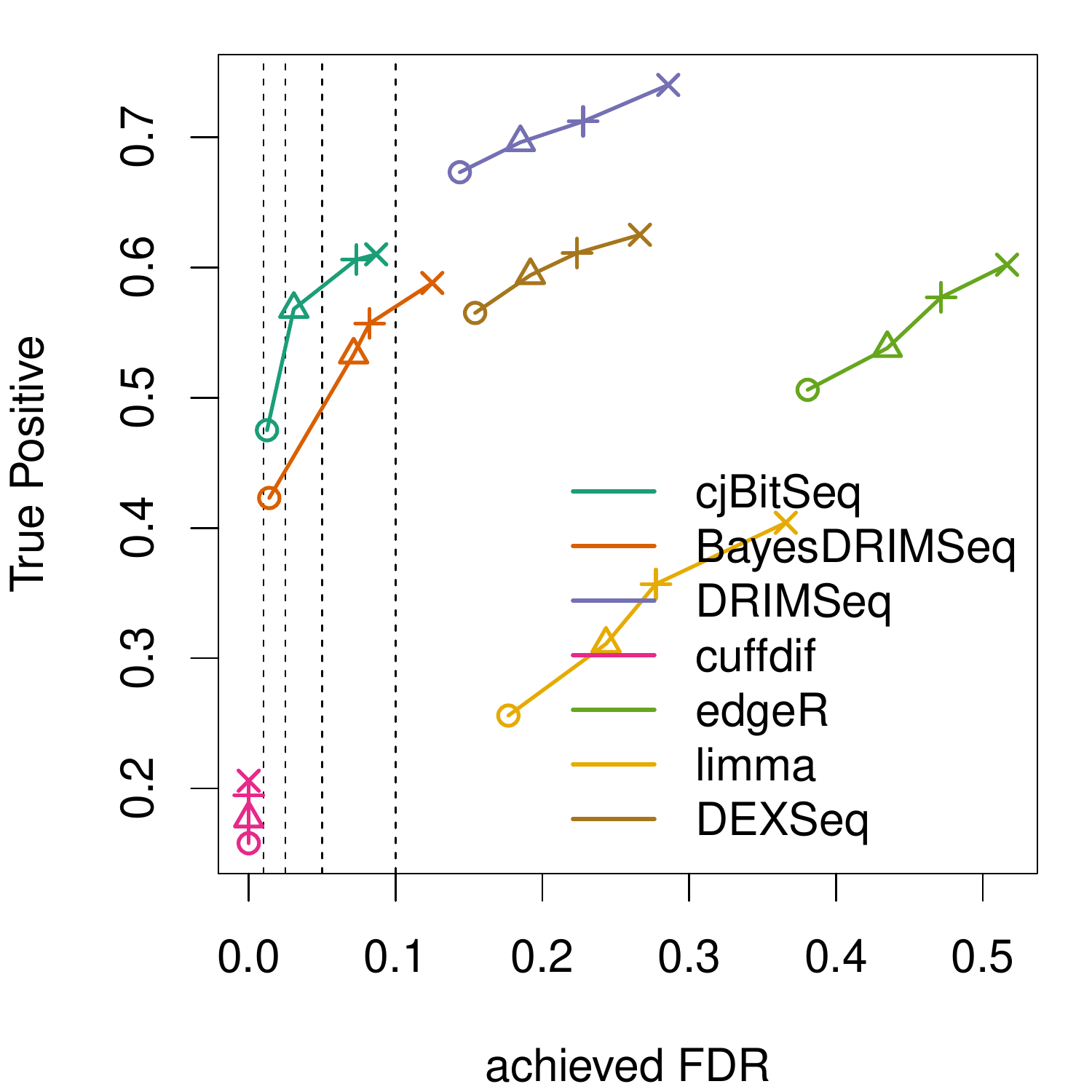}&
\includegraphics[scale=0.30]{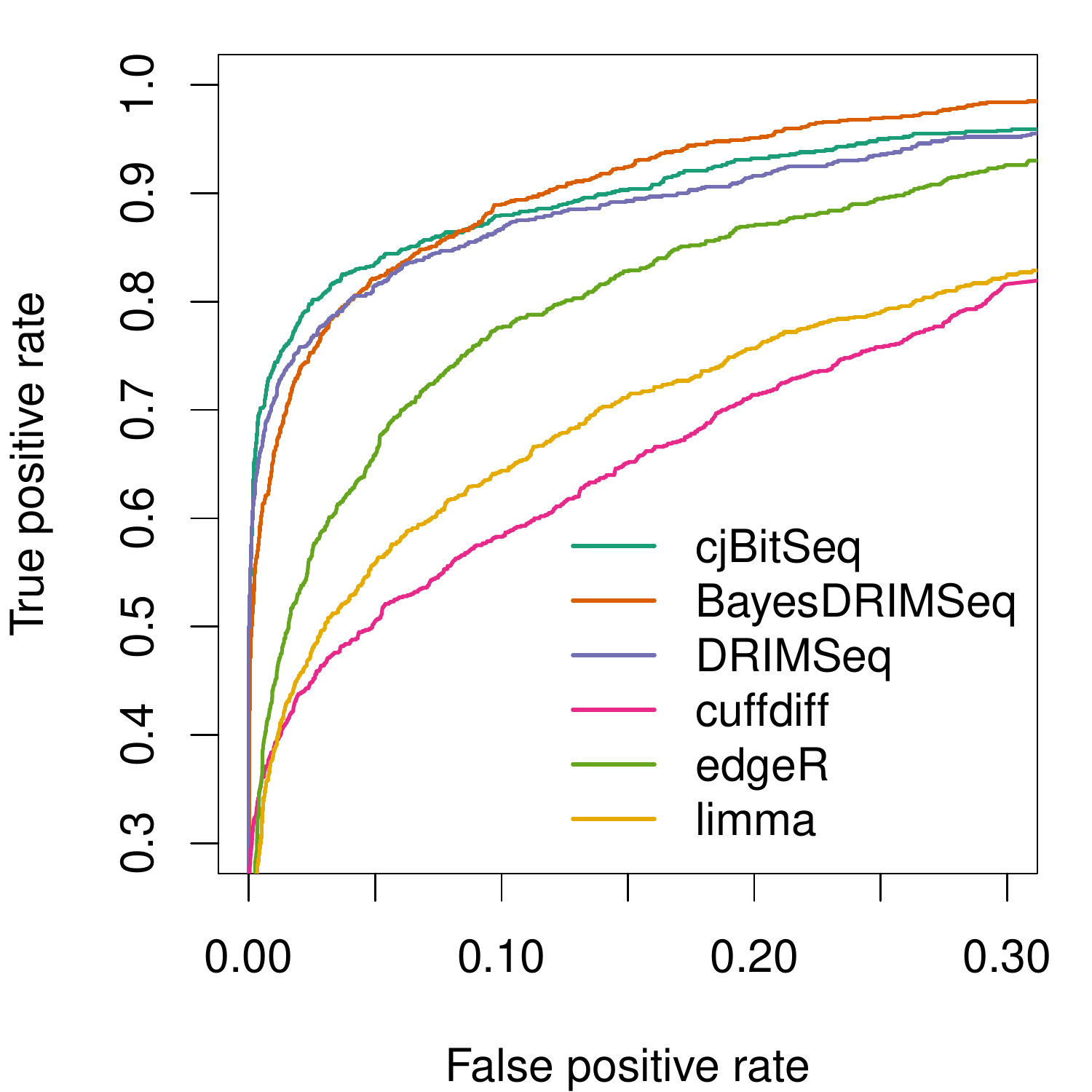}&
\includegraphics[scale=0.30]{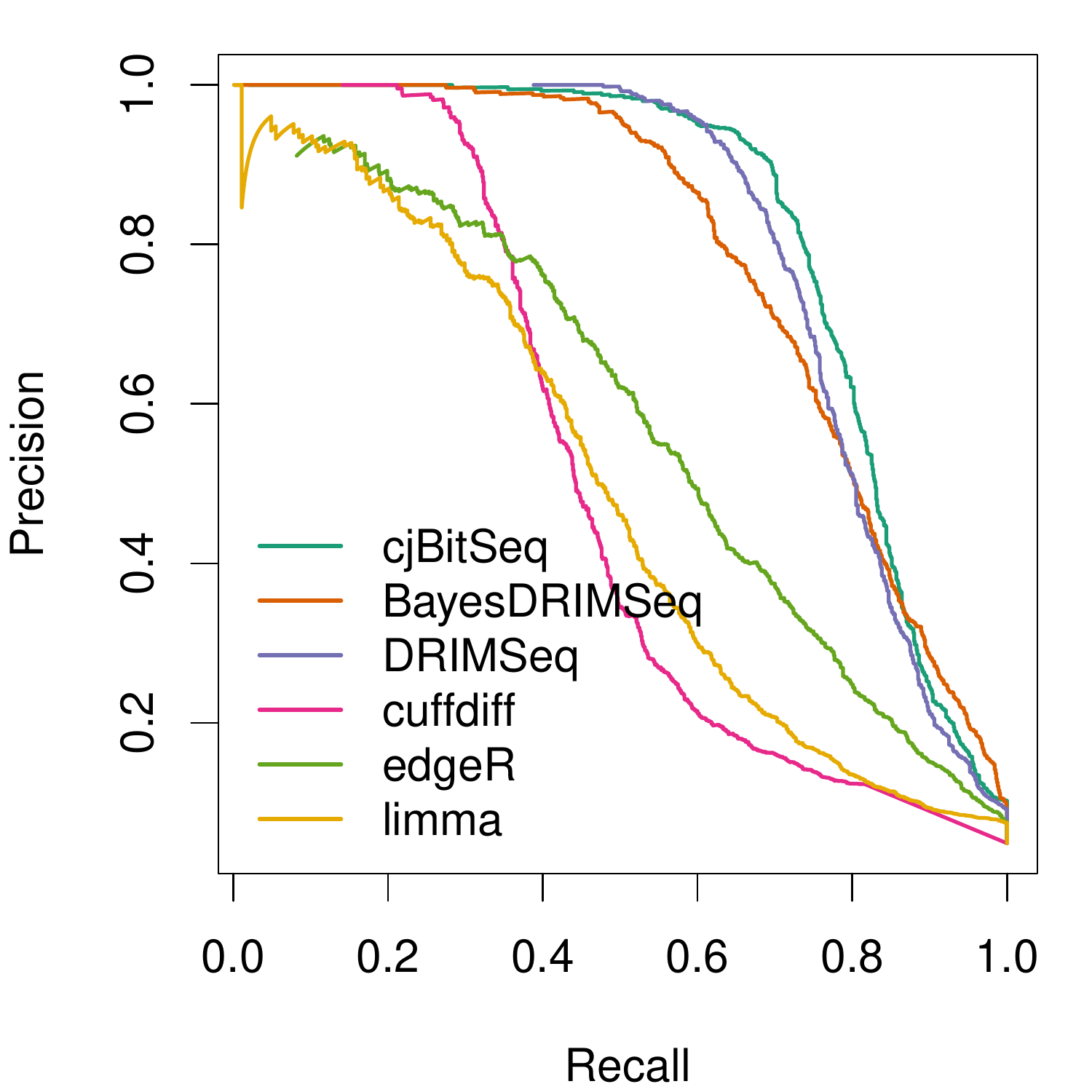}\\
\includegraphics[scale=0.30]{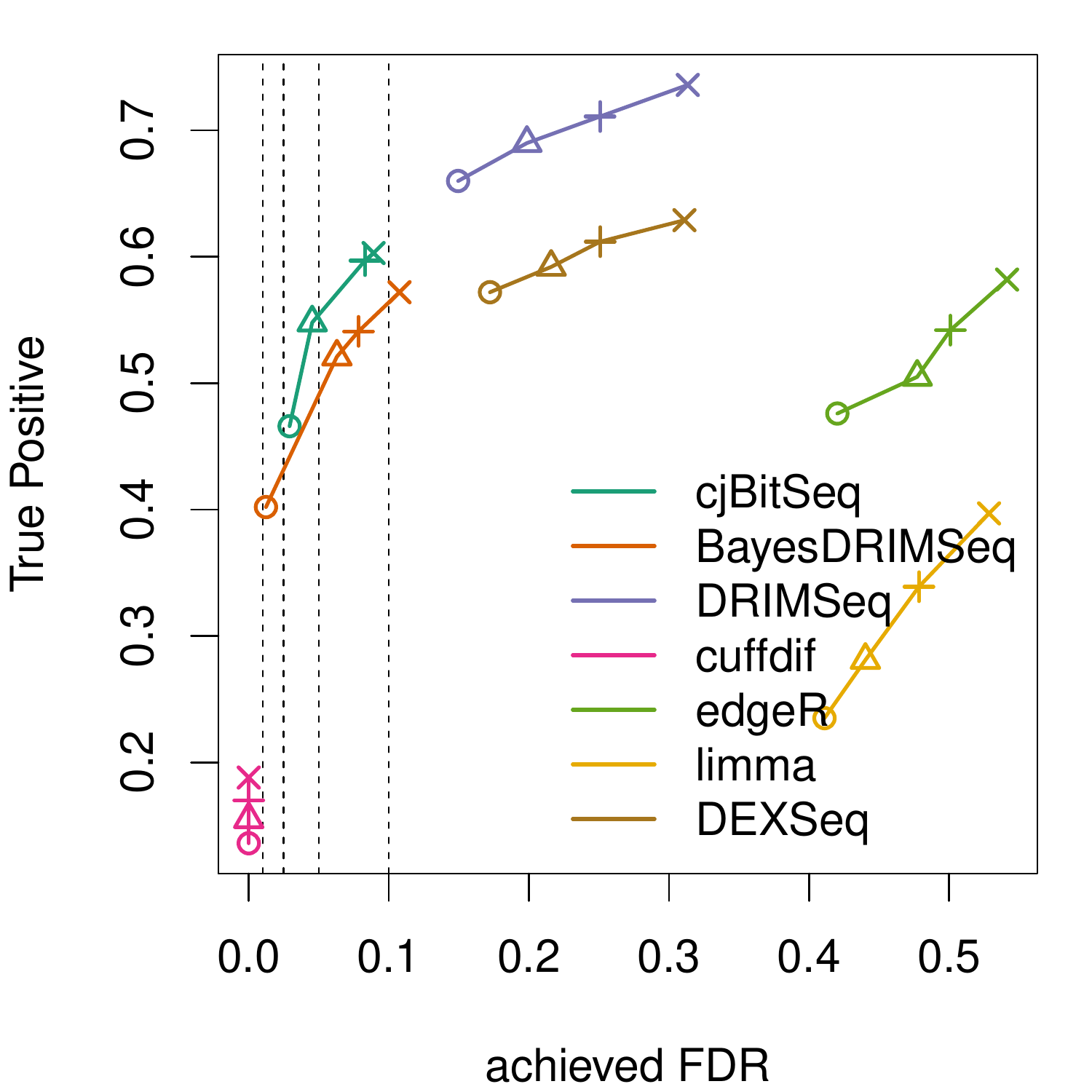}&
\includegraphics[scale=0.30]{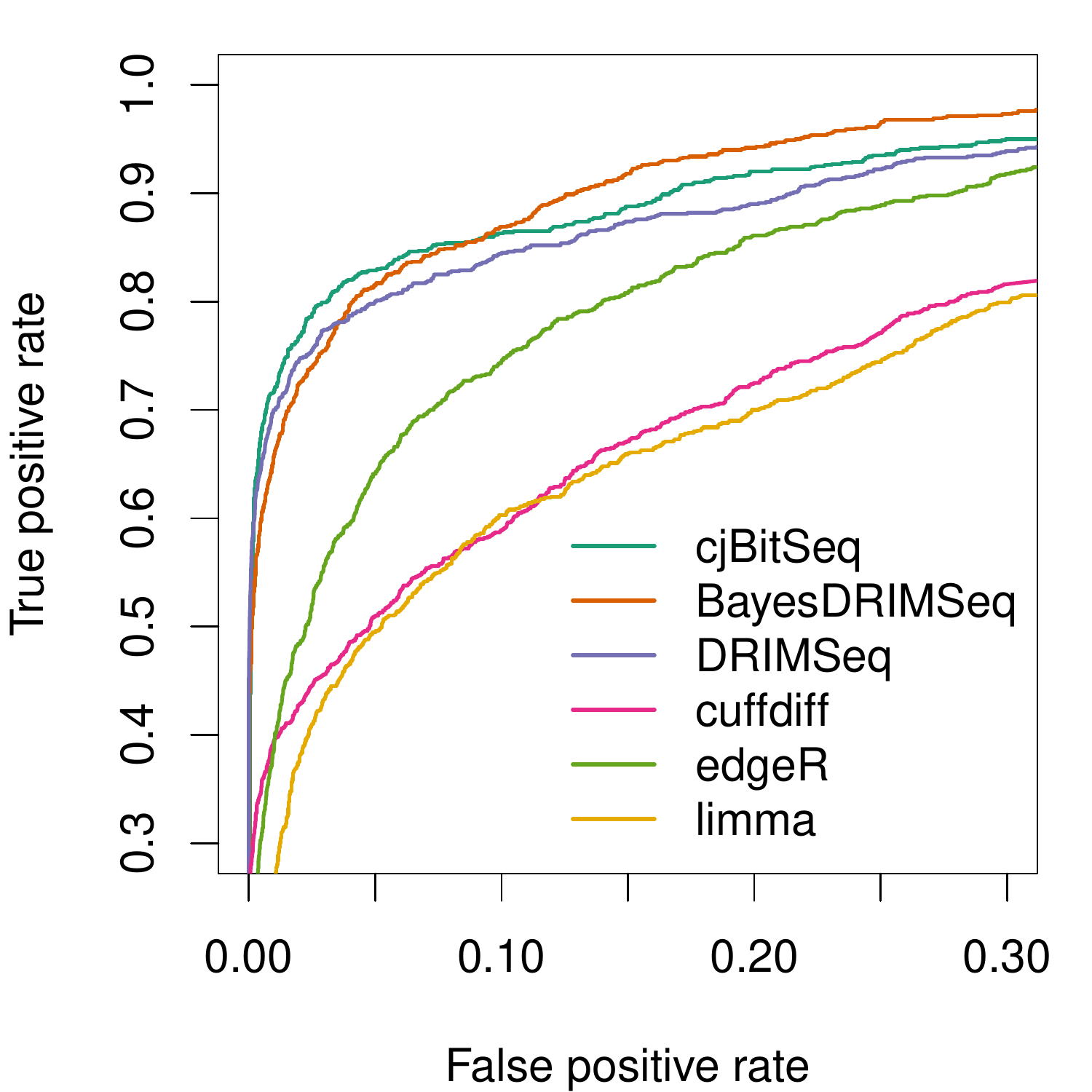}&
\includegraphics[scale=0.30]{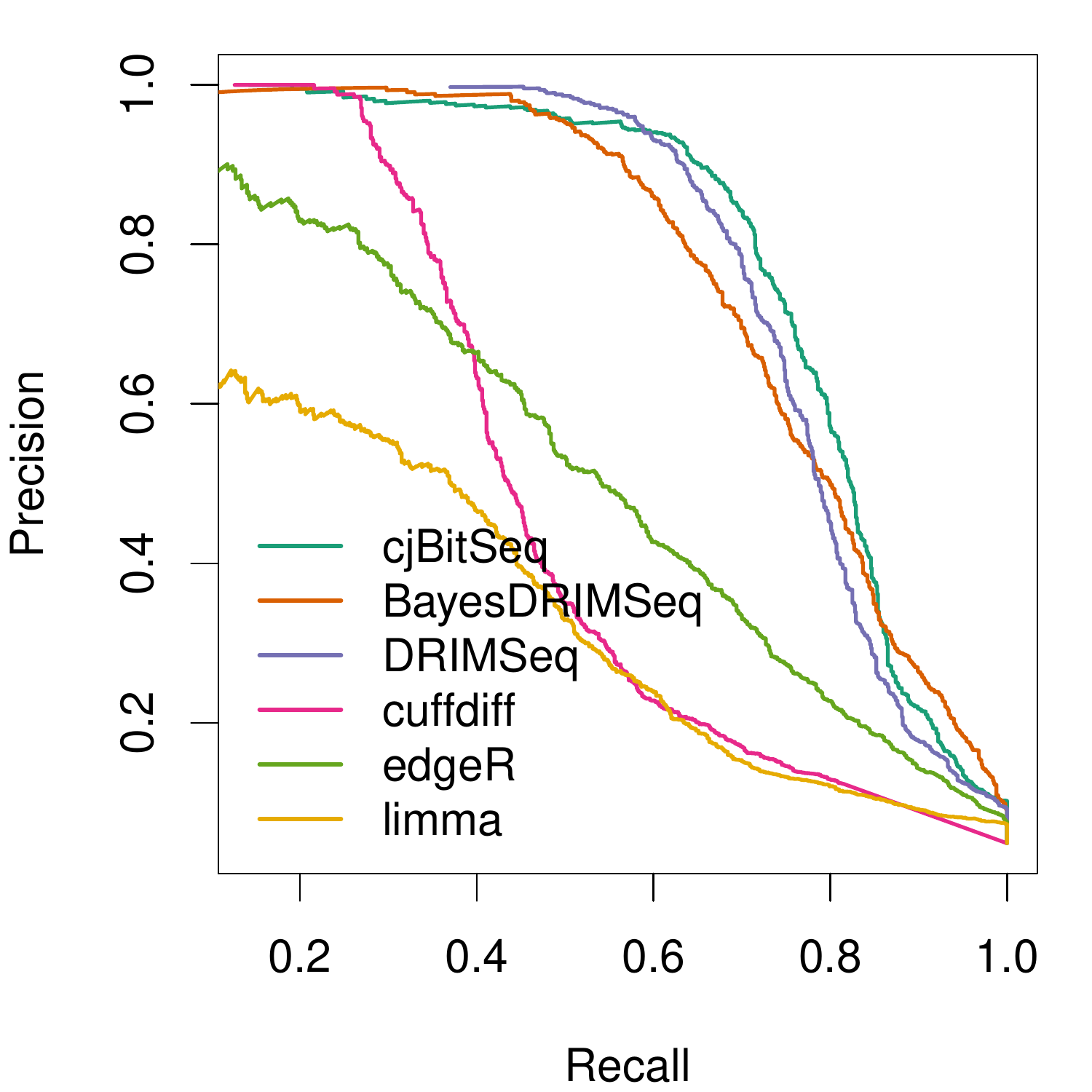}\\
(a) & (b) & (c)
\end{tabular}
\caption{Performance measures for drosophila (1st row), human without DTE (2nd row) and human with DTE (3rd row). (a): power versus achieved FDR plot. The vertical dashed lines show the expected FDR level (0.01, 0.025,0.05,0.1). (b): ROC curve. (c): Precision/recall curve.}
\label{fig:performance}
\end{figure}

For cuffdiff, DRIMSeq, edgeR and limma we use the gene-level p-values at the ROC and precision/recall plots and the adjusted q-values at the power versus achieved FDR plot. However, dexSeq reports only the adjusted q-values, hence this method is not shown at ROC and precision/recall curves. Note that for all these methods, the adjusted q-values correspond to the \cite{benjamini1995} FDR control procedure. For cjBitSeq we used the raw FDR rate \eqref{eq:fdrRaw} at the ROC and precision/recall curves and the adjusted FDR \eqref{eq:fdr} at the power versus achieved FDR plots. For BayesDRIMSeq we used the raw FDR rate \eqref{eq:fdrRaw} at all plots, after pre-filtering isoforms with an average number of reads less than 20.

The performance measures of the evaluated methods are shown in Figure \ref{fig:performance}. Comparing results for the two organisms, it is clear that edgeR, limma, dexSeq and frequentist DRIMSeq exhibit large differences in their ability to control the FDR. In particular, these methods exhibit significantly larger False Discovery rates for the human datasets compared to drosophila. On the other hand, cjBitSeq and BayesDRIMSeq are able to produce consistent results in all cases, being able at the same time to control the FDR within the $(0,0.1)$ area. 

More specifically, for the drosophila example observe that cjBitSeq exhibits smaller achieved FDR rate and larger True Positive Rate compared to DRIMSeq, dexSeq and edgeR. BayesDRIMSeq achieves even smaller FDR rates but the number of True Positives is reduced compared to cjBitSeq. For the human examples we conclude that cjBitSeq exhibit almost similar performance in terms of FDR control, however the former is able to discover a larger number of DTU genes in both cases. DRIMSeq and dexSeq achieve FDR rates between $(0.12, 0.30)$ but DRIMSeq also achieves larger True Positive Rates compared to dexSeq. Cuffdiff exhibits an almost perfect control of the FDR, at the cost of substantially reduced power. The ROC and precision/recall curves, shown at Figure \ref{fig:performance}.(b) and (c) respectively, suggest that cjBitSeq and DRIMSeq are consistently ranked higher than other methods. Overall, we conclude that cjBitSeq outperforms all other methods.

\begin{figure}[t]
\centering
\begin{tabular}{c}
\includegraphics[scale=0.32]{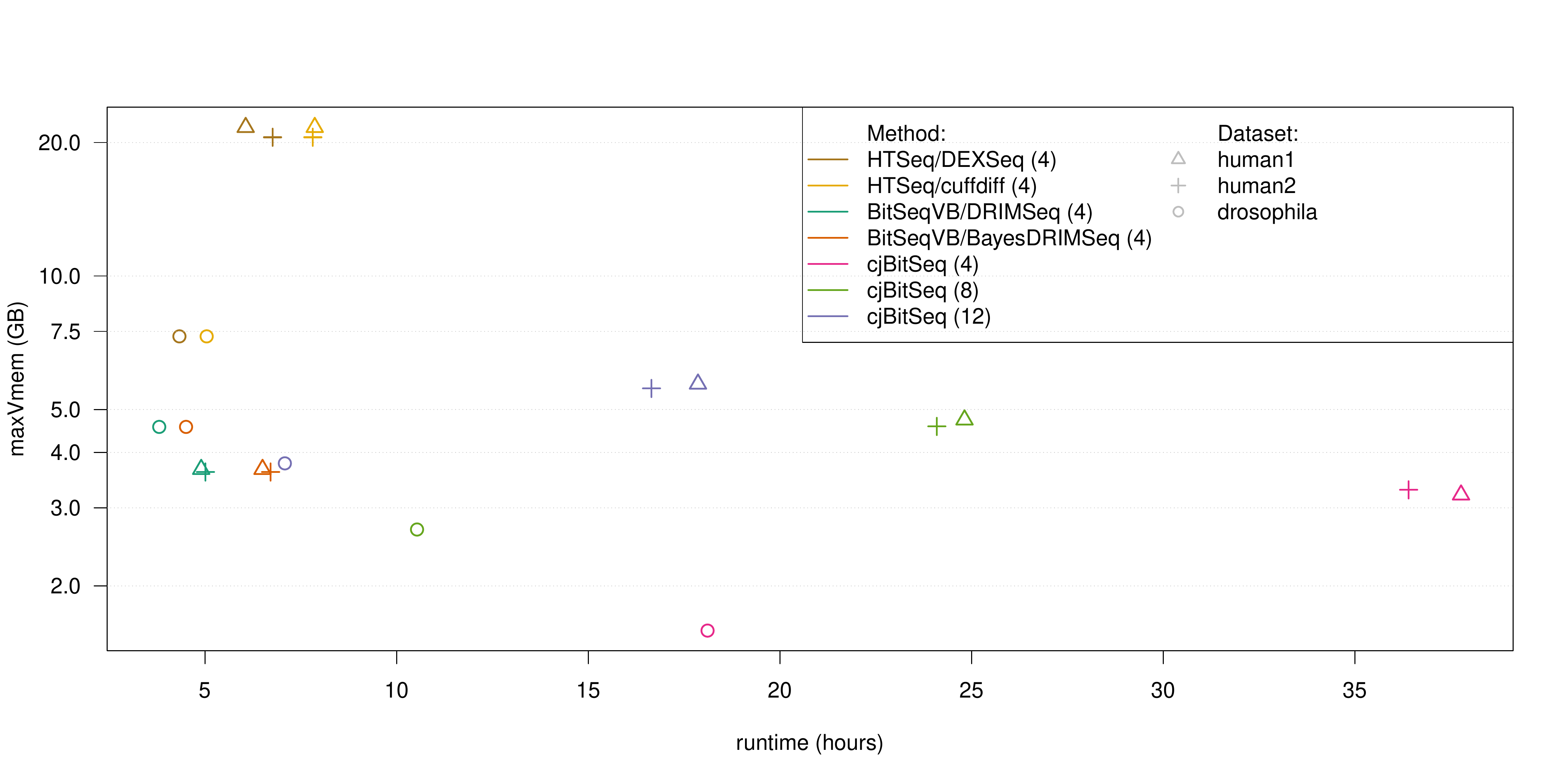}
\end{tabular}
\caption{Wall clock runtime versus maximum value (in log-scale) of virtual memory used. The number of cores used by each process is shown in parenthesis. For each dataset the total number of reads is equal to 150 millions.}
\label{fig:time}
\end{figure}

The run-time per method is illustrated in Figure \ref{fig:time}, with respect to the maximum amount of virtual memory used by each process. For the counting-based methods, the main computational burden of the two-stage pipeline is due to the first stage (that is, either HTSeq or BitSeqVB). DRIMSeq, edgeR, limma which used BitSeqVB as input exhibit nearly identical computing performance so only DRIMSeq is shown. Compared to the counting-based methods, cjBitSeq requires longer computing times, which should be expected given that cjBitSeq performs MCMC sampling on the space of all possible configurations of each transcript using as input the read alignments. However, note that cjBitSeq is quite efficient with respect to the memory used and that both memory and computing time vigorously scale with the number of available cores. Therefore, it is suggested to run cjBitSeq using at least 8 cores, since the memory requirements stay within reasonable levels. Finally, we mention that isoform pre-filtering is also essential for the computing time of BayesDRIMSeq. In case where no filtering takes place, the wallclock time is increased almost $2.5$ times for drosophila and $4.3$ times for the human datasets.

\section{Adenocarcinoma dataset}\label{sec:real}

\begin{figure}[p]
\centering
\begin{tabular}{c}
\includegraphics[scale=0.5]{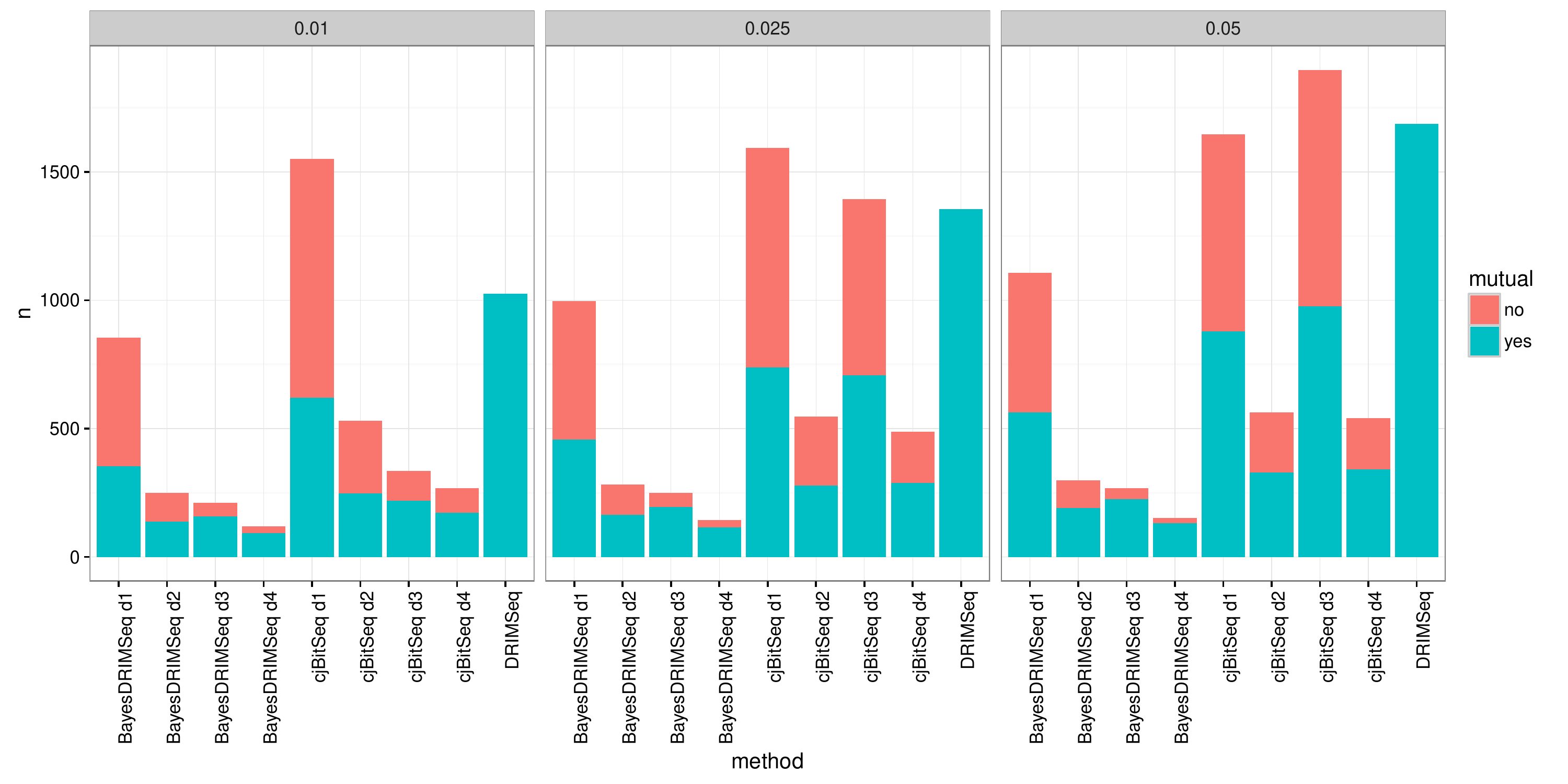}\\
(a) 6 normal versus 6 cancer samples \\
\includegraphics[scale=0.5]{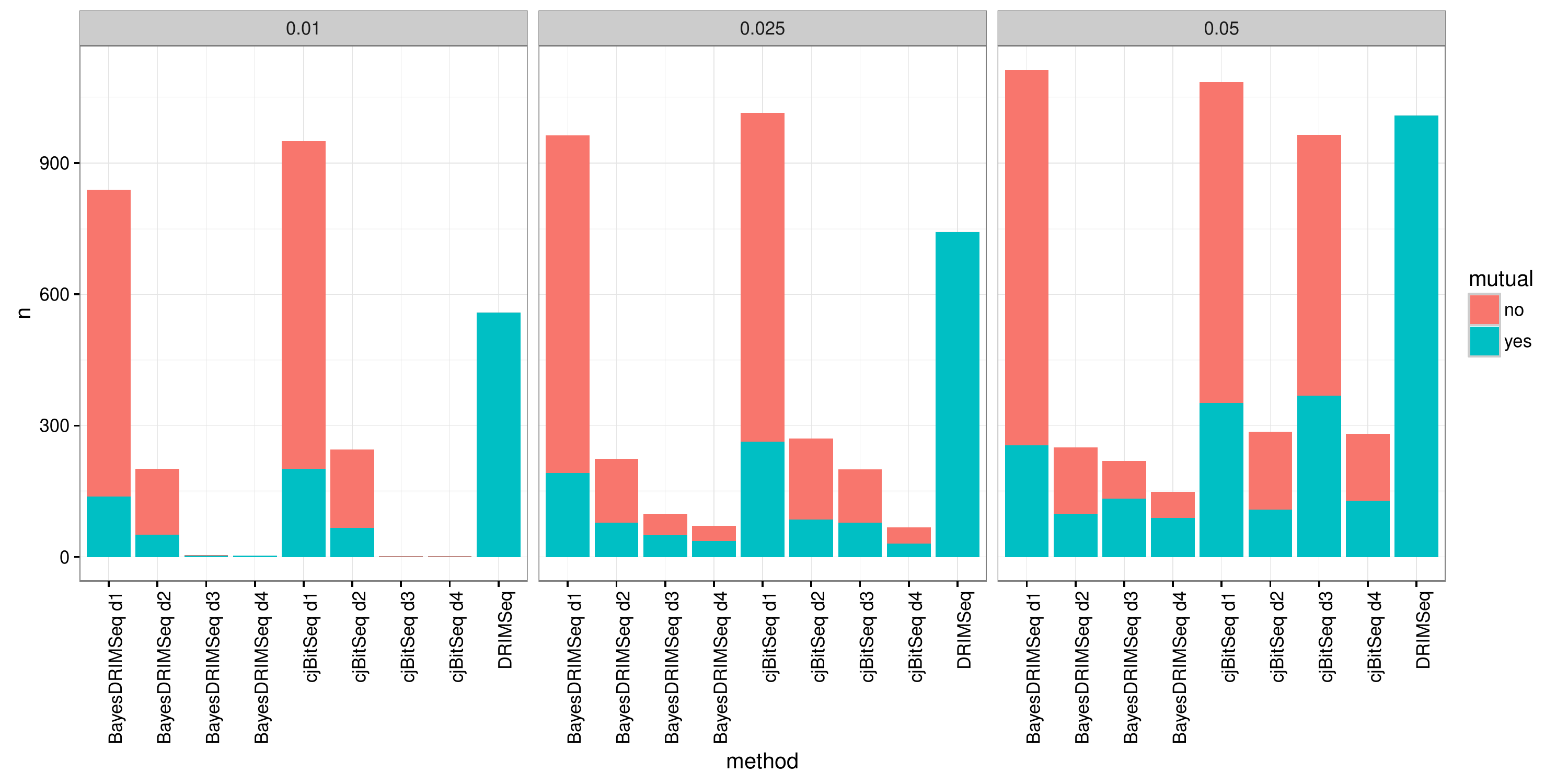}\\
(b) 3 normal versus 3 normal samples
\end{tabular}
\caption{Inferred number of genes with DTU ($n$) at level $\alpha\in\{0.01, 0.025, 0.05\}$ for the comparison of 6 control and 6 tumor samples and null comparisons of 3 versus 3 control samples. For the null comparisons no differential splicing is expected. For the Bayesian methods cjBitSeq and BayesDRIMSeq all 4 decision rules are used. Green color corresponds to the number of DTU genes detected by each method that overlap with DRIMSeq and red corresponds to the opposite case.}
\label{fig:real}
\end{figure}

In this section we benchmark the new Bayesian methods against DRIMSeq using real RNA-seq data from human lung normal and adenocarcinoma samples from six Korean female nonsmoking patients \citep{kim2013high}. The data corresponds to samples from GSM927308 to GSM927319 and was downloaded from 
 NCBI’s Gene Expression Omnibus (GEO) under
the accession number GSE37764: SRR493937, SRR493939, SRR493941, SRR493943, SRR493945, SRR493947,  SRR493949, SRR493951, SRR493953, SRR493955, SRR493957, SRR493959. 

The data consist of paired-end reads with length equal to 78 base pairs which were mapped to the reference transcriptome using Bowtie2. The overall alignment rates and the total number of mapped reads range between $(70\%, 85\%)$ and $(22\times 10^6, 30\times 10^6)$, respectively. Next, BitSeq was used in order to calculate the matrix of alignment probabilities (as input to cjBitSeq) as well as to obtain a matrix of estimated counts per transcript (as input to DRIMSeq and BayesDRIMSeq). 

Following \cite{drimseq2}, we benchmark our methods using two comparisons: (a) a two-group comparison of 6 normal versus 6 cancer samples and (b) ``mock'' comparisons where 3 versus 3 samples from the normal condition are compared. For the latter scenario the expectation is to detect no DTU since replicates of the same condition are compared, although the biological variation between the replicates of the normal condition is high \citep[as noted by][]{drimseq2}. The results are displayed in Figure \ref{fig:real}, using different cutoff values for controlling the FDR. For the 6 normal versus 6 cancer samples comparison (Figure \ref{fig:real}.a), we conclude that all decision rules contain a large amount of genes which overlap with DRIMSeq (green colored regions), especially for the trust-region adjusted rules $d_2$ and $d_4$. For the ``mock'' comparison (Figure \ref{fig:real}.b), at first note that a smaller number of DTU genes is inferred. Second, observe that the decision rule $d_4$ is capable of substantially reducing the number of false discoveries compared to DRIMSeq and that this number is almost zero when using $\alpha = 0.01$.

\section{Discussion}
In this study we exemplified the use of Bayesian methods for inferring genes with differential transcript usage. For this purpose two previously introduced models were modified and extended: cjBitSeq and a Bayesian version of DRIMSeq. After defining proper decision rules we concluded that both methods exhibit superior or comparable performance with other methods. This was achieved by using the decision rule defined in Equation \eqref{eq:fdrRaw}, shown in the ROC and precision-recall curves. According to \eqref{eq:fdrRaw}, the whole sequence of posterior probabilities is transformed with respect to the ordering of the magnitude change of relative expression between conditions. For the read-based method (cjBitSeq) FDR control is improved when the decision rule is combined with a trust region. For the count-based method (BayesDRIMSeq) FDR control is mainly affected by the filtering of low-expressed transcripts, as previously reported under a frequentist context by \cite{Soneson025387}. BayesDRIMSeq exhibits slightly better FDR control than cjBitSeq for the drosophila dataset, however this effect is not so evident for the human datasets. In all cases cjBitSeq is more powerful than BayesDRIMSeq, but at the cost of increased computing time. 

Regarding the analysis of real RNA-seq data, we compared our findings to DRIMSeq. We reported results based on a comparison of two different conditions, as well as ``mock'' comparisons of replicates within the same condition where no evidence of differential expression is expected. We concluded that our DTU lists contain a large number of genes also detected by DRIMSeq. Moreover, using conservative decision rules like $d_4$ we are able to substantially reduce the number of false discoveries when performing comparisons within the same condition.

The methods are freely available from {\tt https://github.com/mqbssppe/cjBitSeq} (cjBitSeq) and {\tt https://github.com/mqbssppe/BayesDRIMSeq} (BayesDRIMSeq). The source code for generating the simulated datasets of \cite{Soneson025387} is available from {\tt https://github.com/\\markrobinsonuzh/diff\_splice\_paper}.

\section*{Acknowledgements}

The research was supported by MRC award MR/M02010X/1, BBSRC award BB/J009415/1 and EU FP7 project RADIANT (grant 305626). The authors would like to acknowledge the assistance given by IT Services and the use of the Computational Shared Facility at The University of Manchester.  Regarding BayesDRIMSeq and replication of simulations, helpful discussions with Mark Robinson, Malgorzata Nowicka and Charlotte Soneson (Institute of Molecular Life Sciences, University of Zurich) are gratefully acknowledged.  

\appendix
\section{Prior Sensitivity of BayesDRIMSeq}

\begin{figure}[t]
\centering
\begin{tabular}{ccc}
\includegraphics[scale=0.30]{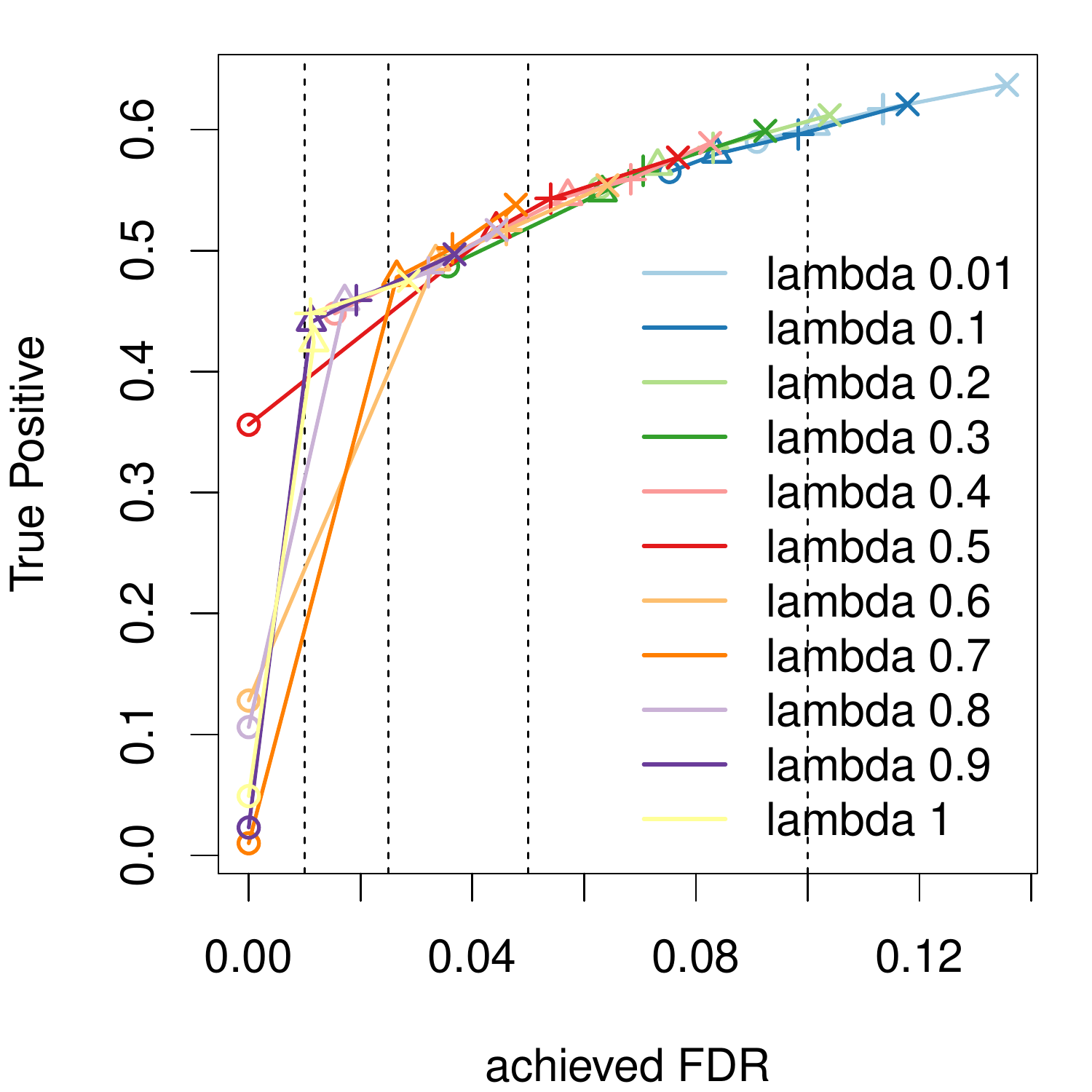}&
\includegraphics[scale=0.30]{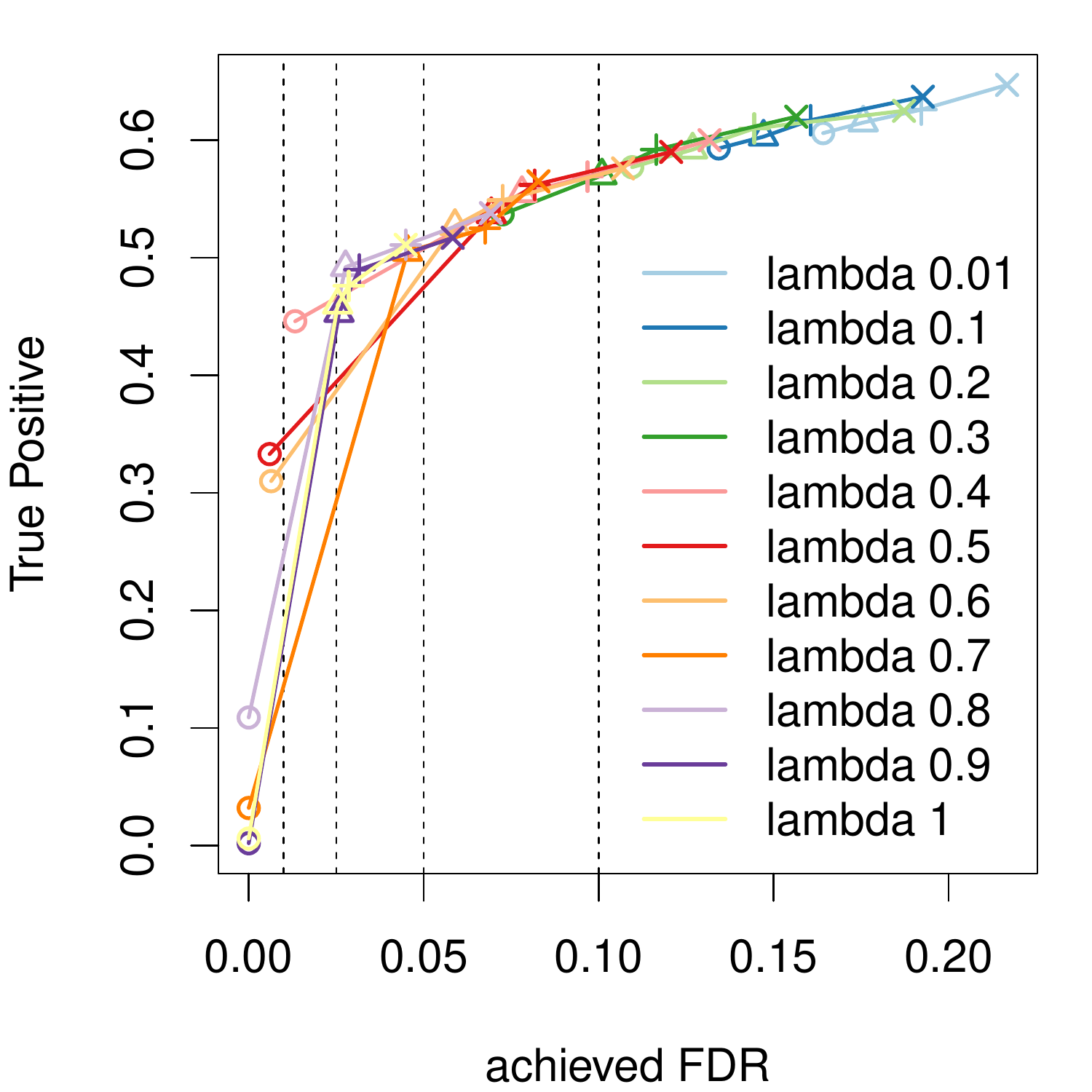}&
\includegraphics[scale=0.30]{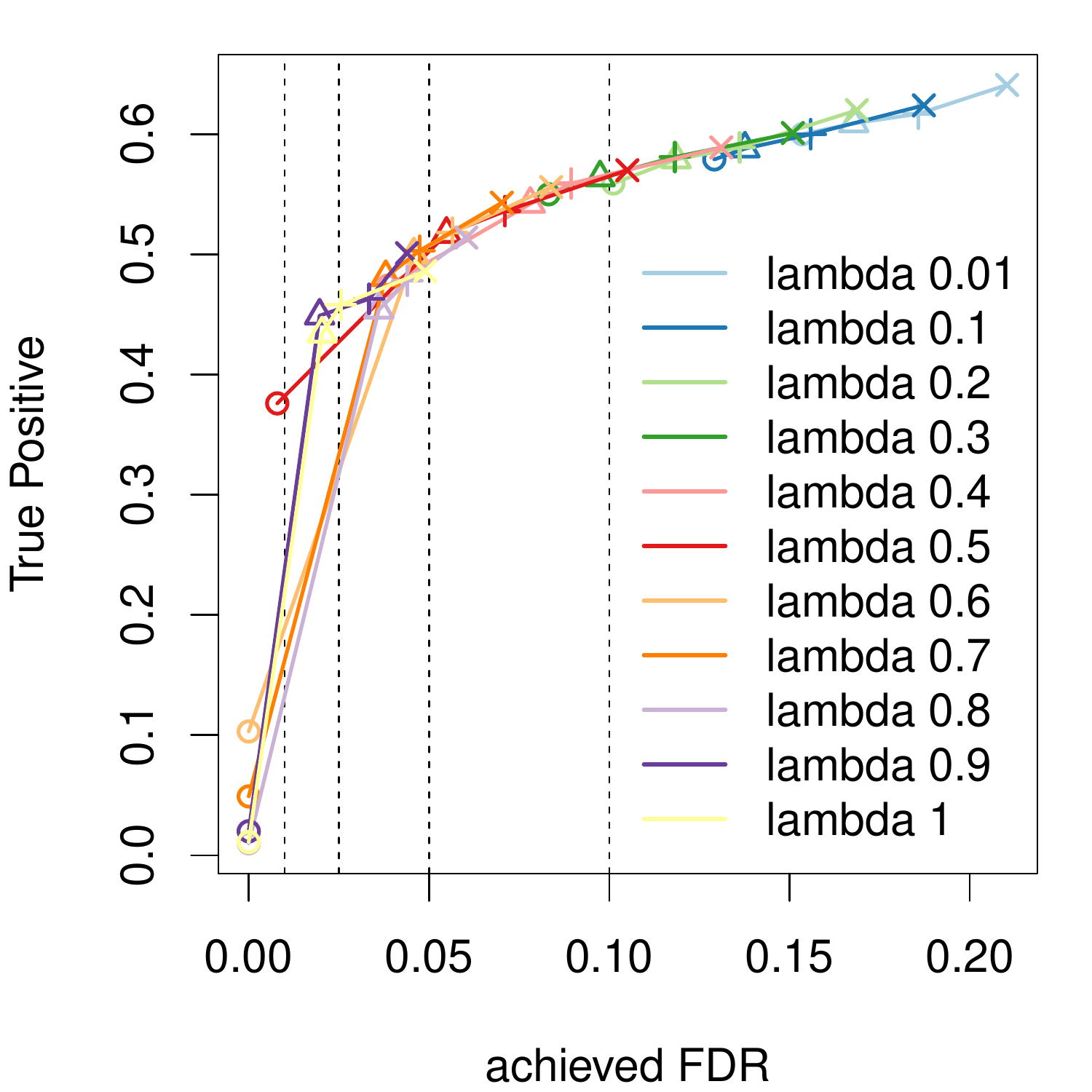}\\
(a) & (b) & (c)
\end{tabular}
\caption{Prior sensitivity of BayesDRIMSeq with respect to $\lambda$.}
\label{fig:laplacePrior}
\end{figure}

According to Equation \eqref{eq:exponential}, the prior assumptions of BayesDRIMSeq are depending on the fixed hyperparameter $\lambda$. Figure \ref{fig:laplacePrior} displays the power versus achieved FDR curves based on the decision $d_3$ as a function of $\lambda \in\{0.01,0.1,0.2,\ldots,1\}$ (after isoform pre-filtering). We conclude that the value $\lambda = 0.5$ offers, perhaps, the best trade-off between power and FDR control. In particular, we note that values smaller than $0.5$ tend to have small power and, on the other hand, values larger than $0.5$ have larger rates of False Discoveries. All results presented in the main paper correspond to $\lambda = 0.5$.

\section{Using Kallisto counts}

\begin{figure}[h]
\centering
\begin{tabular}{c}
\includegraphics[scale=0.4]{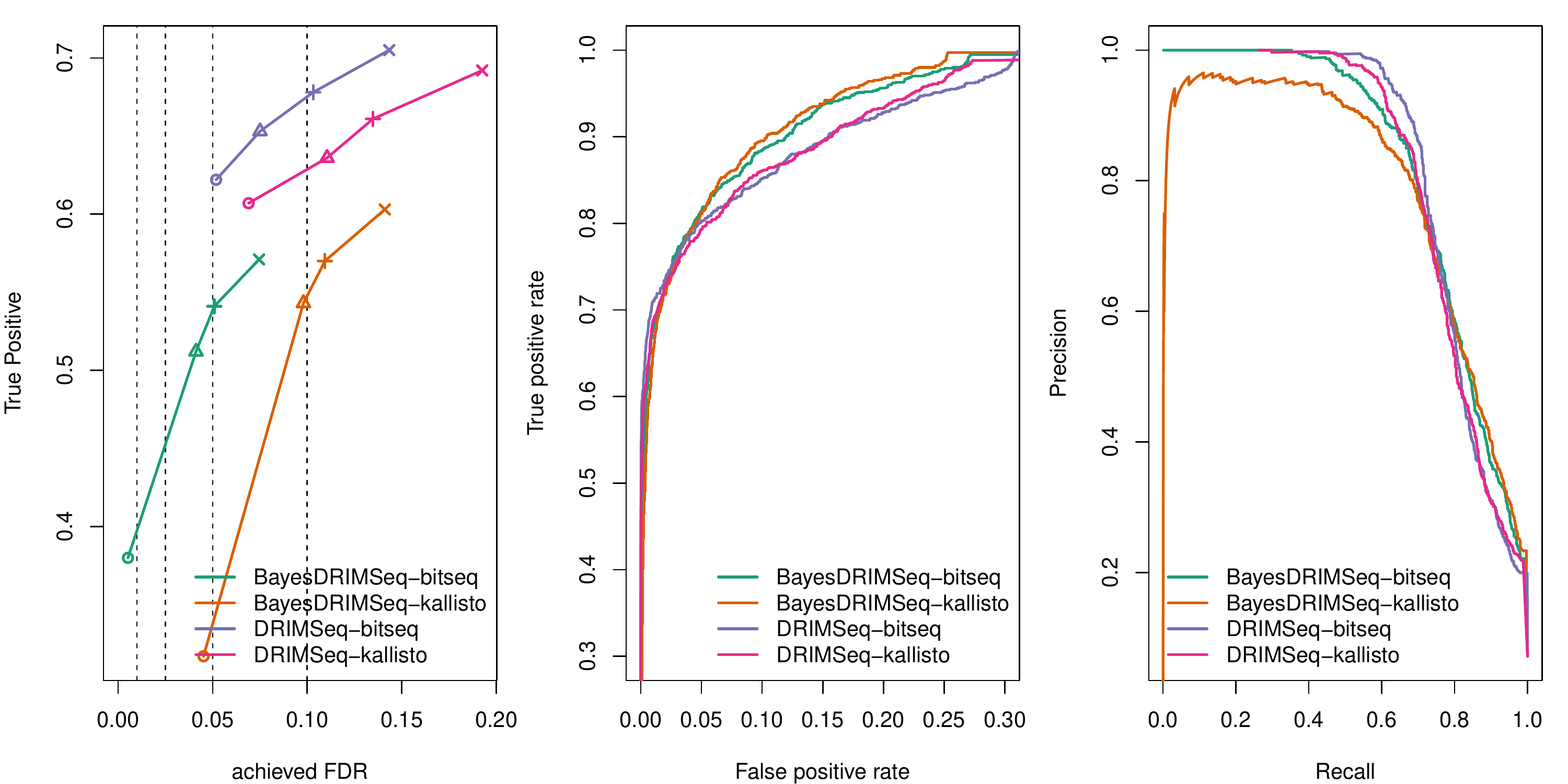}\\
\includegraphics[scale=0.4]{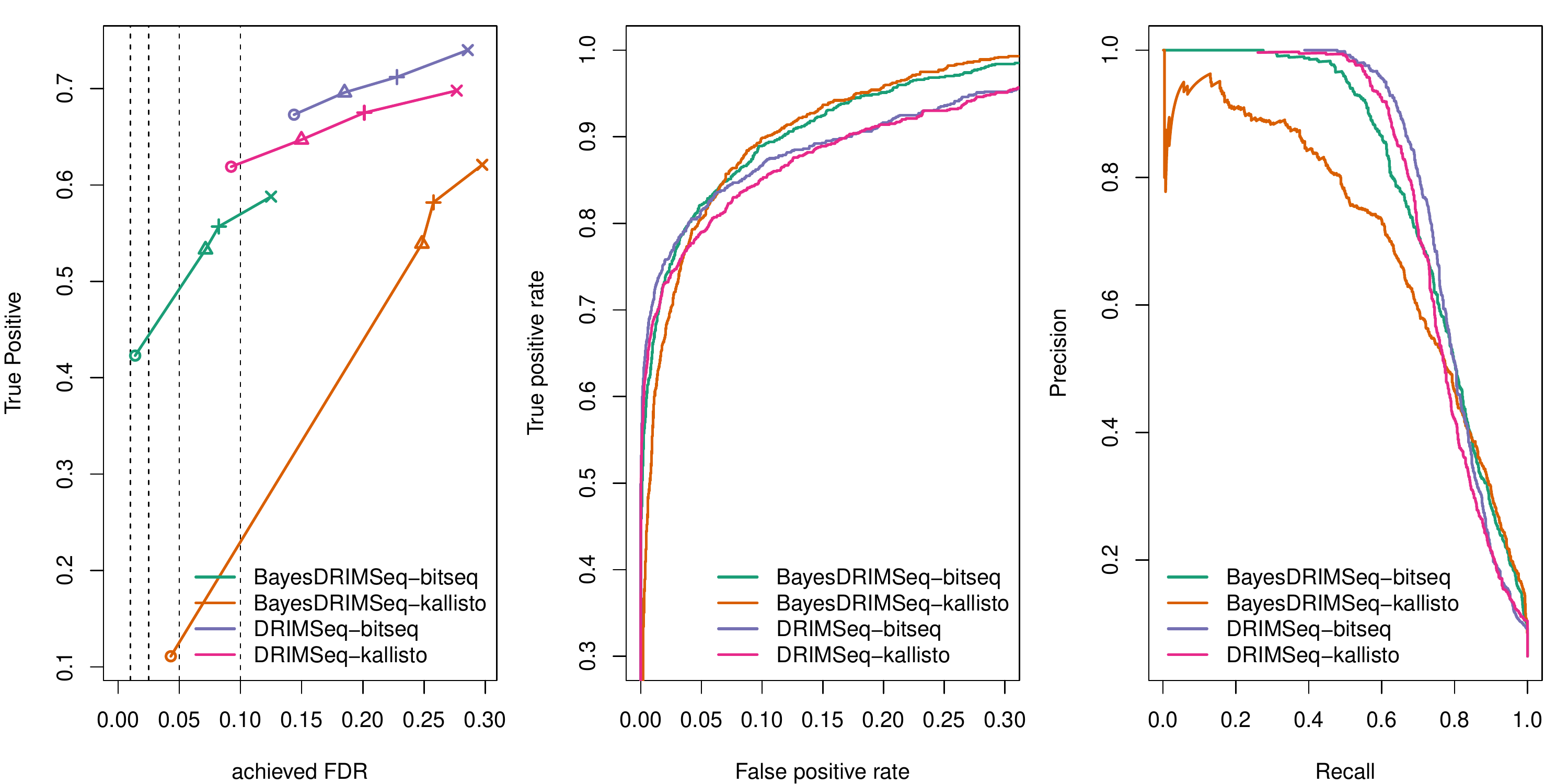}\\
\includegraphics[scale=0.4]{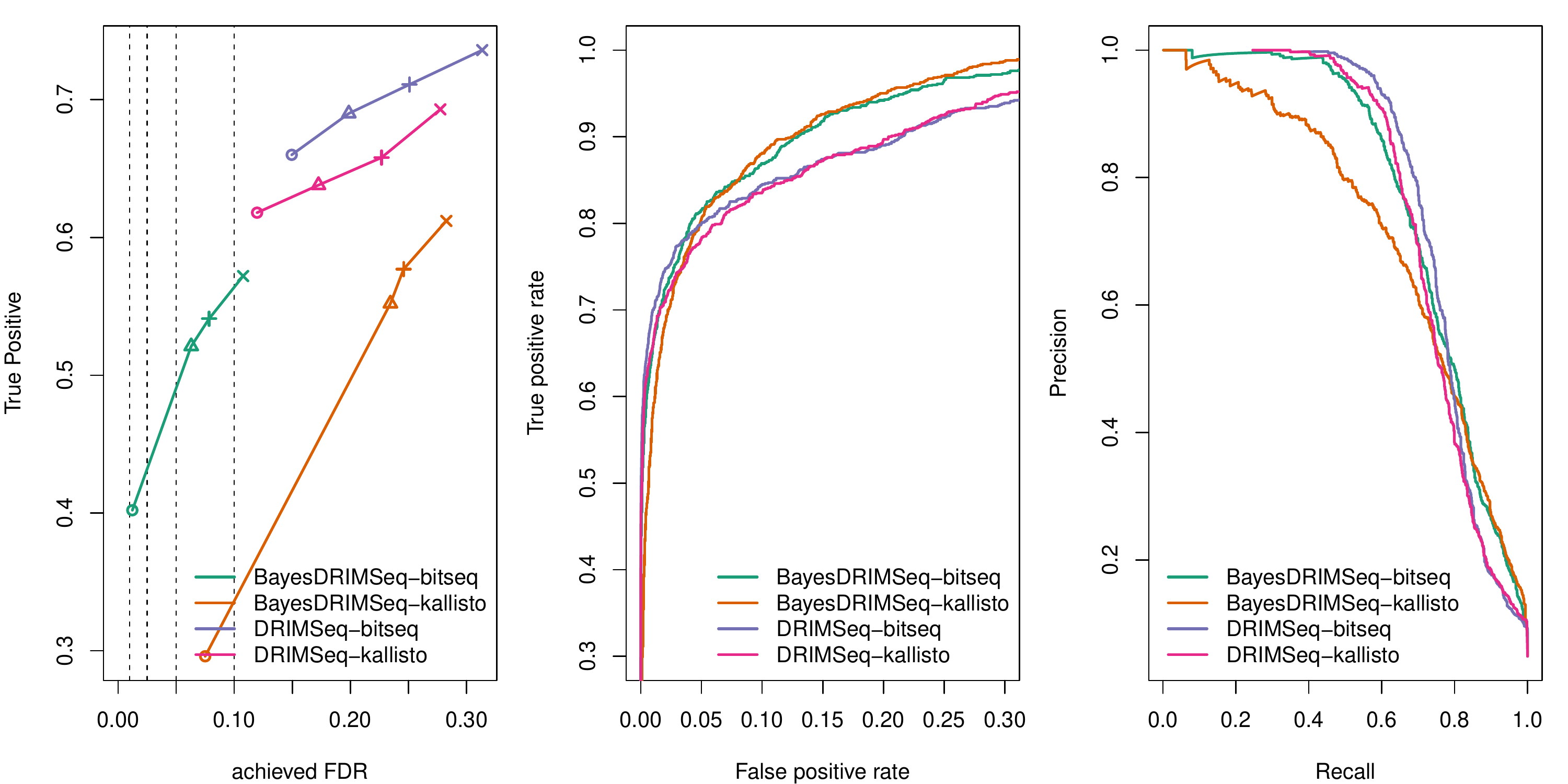}
\end{tabular}
\caption{Comparison of DRIMSeq and BayesDRIMSeq using BitSeq and Kallisto counts for drosophila (first row) and human data (second and third row).}
\label{fig:kallisto}
\end{figure}

In the main text we used BitSeqVB count estimates as input to DRIMSeq and BayesDRIMSeq. According to the recent study of \cite{bitseqVB}, BitSeqVB is ranked as one of the most accurate methods for estimating transcript expression levels. Since there is a variety of alternative methods for this purpose, we compare the performance when Kallisto \citep{kallisto} counts are being used as input. As  shown in Figure \ref{fig:kallisto}, we conclude that in drosophila data both BayesDRIMSeq and DRIMSeq perform better when BitSeqVB counts are used. However there is no clear ordering in the human datasets: in both cases BitSeqVB counts correspond to increased power but at the cost of slightly worse FDR calibration. Finally, ROC and precision-recall curves suggest that BitSeqVB leads to slightly increased performance for both methods. 

\bibliographystyle{abbrvnat}
\bibliography{main}

\end{document}